\documentclass[aps,pra,twocolumn,groupedaddress,showpacs]{revtex4}

\usepackage{graphicx}
\usepackage{amsmath}
\usepackage{bm}% bold math

\begin{document}

%\preprint{}

\title{Creating vortons and three-dimensional skyrmions from domain wall annihilation 
with stretched vortices in Bose-Einstein condensates}


\author{Muneto Nitta$^{1}$}
\author{Kenichi Kasamatsu$^{2}$}
%\email[]{kenichi@phys.kindai.ac.jp}
\author{Makoto Tsubota$^{3}$}
\author{Hiromitsu Takeuchi$^{4}$}
%\thanks{}
\affiliation{
$^1$Department of Physics, and Research and Education Center for Natural 
Sciences, Keio University, Hiyoshi 4-1-1, Yokohama, Kanagawa 223-8521, Japan\\
$^2$Department of Physics, Kinki University, Higashi-Osaka, 577-8502, Japan \\
$^3$Department of Physics, Osaka City University, Sumiyoshi-Ku, Osaka 558-8585, Japan \\
$^4$Graduate School of Integrated Arts and Sciences, Hiroshima
University, Kagamiyama 1-7-1, Higashi-Hiroshima 739-8521, Japan
}


\date{\today}
\begin{abstract}
We study a mechanism to create a vorton 
or three-dimensional skyrmion in 
phase-separated two-component BECs 
with the order parameters $\Psi_1$ and $\Psi_2$ 
of the two condensates.  
We consider a pair of a domain wall (brane) 
and an anti-domain wall (anti-brane)
stretched by vortices (strings), 
where the $\Psi_2$ component with 
a vortex winding is sandwiched by 
two domains of the $\Psi_1$ component. 
The vortons appear when the domain wall pair annihilates. 
Experimentally, this can be realized by preparing 
the phase separation in the order $\Psi_1$, $\Psi_2$ and $\Psi_1$ components, 
where the nodal plane of a dark soliton in $\Psi_1$ component is 
filled with the $\Psi_2$ component with vorticity. 
By selectively removing the filling $\Psi_2$ component gradually 
with a resonant laser beam, the collision of the brane and anti-brane 
can be made, creating vortons.
\end{abstract}
%\keywords{}

\maketitle

\section{Introduction} \label{Intro}

Quantized vortices are one of remarkable consequences 
of superconductivity and superfluidity. 
In multi-component superfluids and superconductors, 
there appear many kinds of exotic vortices.
When a vortex of one condensate 
traps another condensate inside its core, 
a supercurrent or superflow of the latter 
can exist along the vortex line. 
Such a vortex is called a superconducting 
or superflowing cosmic string in cosmology 
\cite{Witten:1984eb}.
Because of the Meissner effect, 
superconducting strings exclude magnetic fields 
like superconductive wires, 
so that they are proposed to explain several cosmological 
phenomena related to galactic magnetic fields.
When a superconducting string is closed 
and ``twisted", {\it i.e.}, 
when the second condensate inside the string core 
has a non-trivial winding along the string loop, 
the supercurrent persistently flows along the loop 
and makes it stable.
Such a twisted vortex loop is called a ``vorton", a particle-like 
soliton made of a vortex \cite{Davis:1988jq,Radu:2008pp}.
While vortons were discussed in $^3$He superfluids \cite{Volovik}, 
they are considered to be 
a candidate of dark matter, 
and a possible source of ultra high energy cosmic ray. 
There have been a lot of study about 
their stability, interaction, and applications 
to cosmology \cite{Vilenkin:2000}.

On the other hand, three-dimensional (3D) skyrmions 
are topological solitons (textures) characterized by 
the third homotopy group $\pi_3(SU(2))\simeq {\bf Z}$ 
in a pion effective field theory. 
Skyrmions were proposed to be baryons \cite{Skyrme:1961vq}. 
Since their proposal, the skyrmions have been studied 
for a long time about their stability, interaction, 
and applications to nuclear physics \cite{Manton:2004tk}.

Both 3D skyrmions and vortons 
have been fascinating subjects in high energy physics 
and cosmology for decades, and a lot of works have been done already, 
but they have yet to be observed in nature.
On the other hand, these topological excitations,  
3D skyrmions \cite{Ruostekoski:2001fc,Battye:2001ec,3D-Skyrmions2,Savage:2003hh,Ruostekoski:2004pj,Wuster:2005,Oshikawa:2006} 
and vortons \cite{Metlitski:2003gj,Bedaque:2011sb}, 
can be realized in 
Bose--Einstein condensates (BECs) of ultracold atomic gasses.
Moreover 3D skyrmions and vortons 
have been shown to be topologically equivalent 
in two-component BECs \cite{Ruostekoski:2001fc,Battye:2001ec}. 
BECs are extremely 
flexible systems for studying solitons (or topological defects) since optical
techniques can be used to control and directly visualize the condensate 
wave functions \cite{Leslie:2009}. 
Interest in various topological defects in BECs
with multicomponent order parameters has been increasing; the structure, stability, and creation and detection
schemes for monopoles \cite{Stoof,Martikainen,Savage,Pietila:2009},  
knots \cite{Kawaguchi:2008xi}
and non-Abelian vortices \cite{Semenoff:2006vv} 
have been discussed \cite{review}.

It is, however, still unsuccessful to 
create vortons and 3D skyrmions experimentally, 
although the schemes to create and stabilize
them have been theoretically proposed 
\cite{Ruostekoski:2001fc,Battye:2001ec,3D-Skyrmions2,Savage:2003hh,Ruostekoski:2004pj,Wuster:2005,Oshikawa:2006}.  
In the present study, we propose how to 
create vortons or 3D skyrmions in two-component BECs 
from domain walls and quantized vortices. 
Specific examples of the system include a BEC mixture of two-species atoms 
such as $^{87}$Rb--$^{41}$K \cite{Thalhammer} or $^{85}$Rb--$^{87}$Rb \cite{Papp}, 
where the miscibility and immiscibility can be controlled by tuning 
the atom--atom interaction via Feshbach resonances. 
Here, the domain wall is referred to as an interface boundary of 
phase-separated two-component BECs. Although the interface has a finite thickness, the wall 
is well-defined as the plane in which both components have the same amplitude. 
Since a description of two-component BECs can be mapped to 
the $O(3)$ nonlinear sigma model (NL$\sigma$M) by introducing a pseudospin representation of the order 
parameter \cite{Kasamatsu2,Babaev:2001zy,Babaev:2008zd}, 
the resultant wall--vortex composite 
soliton corresponds to the Dirichlet(D)-brane soliton described in 
Refs.~\cite{Gauntlett:2000de,Shifman:2002jm,
Isozumi:2004vg,Eto:2006pg,Eto:2008mf}, 
which resembles a D-brane in string theory 
\cite{Polchinski:1995mt,Leigh,Polchinskibook}. 
Such a D-brane soliton has been already numerically 
constructed by us in two-component BECs \cite{Kasamatsu:2010aq}.
We have found that these composite solitons are energetically stable 
in rotating, trapped BECs and are experimentally feasible 
with realistic parameters. 
Similar configuration has been also studied in 
spinor BECs \cite{Borgh:2012es}.

A brane--antibrane annihilation was demonstrated 
to create some topological defects 
in superfluid $^{3}$He \cite{Bradley}. 
However, a physical explanation of the creation mechanism 
of defects still remains unclear. 
The intriguing experiment that mimicked the brane--antibrane annihilation 
was performed in cold atom systems 
with the order parameters $\Psi_1$ and $\Psi_2$ 
of two-component BECs by Anderson {\it et al}. \cite{Anderson}.
They prepared the configuration of 
the phase separation in the order $\Psi_1$, $\Psi_2$ and $\Psi_1$ components, 
where the nodal plane of a dark soliton in one component was filled 
with the other component. 
By selectively removing the filling component 
with a resonant laser beam, they made a planer dark-soliton in 
a single-component BEC. 
Then, the planer dark soliton in 3D system 
is dynamically unstable for its transverse deformation 
(known as snake instability) \cite{Anderson}, 
which results in the decay of the dark soliton 
into vortex rings. 
In the two-component BECs, we have numerically simulated 
brane--anti-brane annihilations,  
which resulted in vortex loops \cite{Takeuchi:2011}.

In this paper, we consider a junction of  
a D-brane soliton and its anti-soliton, 
namely a pair of a domain wall and an anti-domain wall 
stretched by vortices. 
We give an approximate analytic solution 
for a pair of the D-brane and anti-D-brane 
in the $O(3)$ NL$\sigma$M.
We show that this unstable configuration decays 
into a vorton or a 3D skyrmion, 
instead of an untwisted vortex ring \cite{Anderson} 
in the case without stretched vortices. 
Experimentally, this can be realized by preparing 
the phase separation in the order $\Psi_1$, $\Psi_2$ and $\Psi_1$ 
components, 
and rotating the intermediate $\Psi_2$ component.
By selectively removing the filling $\Psi_2$ component gradually 
with a resonant laser beam, the collision of the D-brane and anti-D-brane 
can be made, to create vortons.

This paper is organized as follows. 
In Sec.~\ref{sec:system}, we present 
the Gross-Pitaevski energy functional of 
two-component BECs, 
and rewrite it in the form of NL$\sigma$M.
In Sec.~\ref{sec:wall-anti-wall}, 
after constructing a phase separation, {\it i.e.}, 
a domain wall configuration 
in NL$\sigma$M, 
we consider a pair of a domain wall and an anti-domain wall.
We discuss a creation of vortex in two dimensions, 
and a creation of vortex loops in three dimensions 
after a pair annihilation of the domain walls.
In Sec.~\ref{sec:brane-anti-brane}, 
we consider a pair of a domain wall and an anti-domain wall 
with vortices stretched between them. 
We show that when a vortex loop encloses 
$n$ of the stretched vortices, the phase of the $\Psi_2$ component 
winds $n$ times, {\it i.e.}, it is a vorton with $n$ twist. 
We also confirm a vorton with $n=1$ is topologically equivalent 
to a 3D skyrmion.
Sec.~\ref{sec:summary} is devoted to a  summary and discussion.

%%%%%%%%%%%%%%%%%%%%%%%%%%%%%
\section{System \label{sec:system}}
\subsection{Gross-Pitaevski energy functional}
The order parameter of two-component BECs is 
\begin{equation}
{\bf \Psi} = (\Psi_{1}, \Psi_{2}),
\end{equation} 
where 
\begin{equation}
\Psi_j=\sqrt{\rho_j}e^{i\theta_{j}} \quad (j=1,2)
\end{equation} 
are the macroscopically occupied spatial 
wave function of the two components with the density $\rho_{j}$ and  the phase $\theta_{j}$. 
The order parameter can be represented by the pseudospin 
\begin{equation}
{\bf s} = (s_{1},s_{2},s_{3}) = (\sin \theta \cos \phi , \sin \theta \sin \phi , \cos \theta ) 
\end{equation}
with a polar angle $\theta = \cos^{-1} [(\rho_{1}-\rho_{2})/\rho]$ 
and an azimuthal angle 
$\phi=\theta_{2}-\theta_{1}$ as 
\begin{equation}
{\bf \Psi}= \sqrt{\rho} e^{i \frac{\Theta}{2}} \left( \cos \frac{\theta}{2} 
e^{-i \frac{\phi }{ 2}},  \sin \frac{\theta}{2} e^{i \frac{\phi }{ 2}} \right), 
\label{peudsospinrep}
\end{equation} 
where $\rho =\rho_{1} + \rho_{2}$ and $\Theta = \theta_{1} + \theta_{2}$ represent the local density 
and phase, respectively \cite{Kasamatsu2}. 

The solutions of the solitonic structure in two-component BECs are given 
by the extreme of the Gross--Pitaevski (GP) energy functional 
\begin{eqnarray}
E [{\bf \Psi}] = \int d^{3} x \biggl\{ \sum_{j = 1,2} 
\biggl[ \frac{\hbar^{2}}{2m_{j}}  \left| \nabla \Psi_{j} \right|^{2} 
 + (V_{j} - \mu_{j}) |\Psi_{j}|^{2} \nonumber \\ 
+ \frac{g_{jj}}{2} |\Psi_{j}|^{4} \biggr] 
+ g_{12} |\Psi_{1}|^{2} |\Psi_{2}|^{2} \biggr\}.  \label{GPene}
\end{eqnarray}
Here, $m_{j}$ is the mass of the $j$th component and $\mu_j$ is its chemical potential. 
The BECs are confined by the harmonic trap potential 
\begin{equation}
V_{j}=\frac{1}{2} m_{j} (\omega_{x}^{2} x^{2}+\omega_{y}^{2} y^{2}+\omega_{z}^{2} z^{2}). 
\end{equation}
The coefficients $g_{11}$, $g_{22}$, and $g_{12}$ represent the 
atom--atom interactions. They are expressed in terms of the 
s-wave scattering lengths $a_{11}$ and $a_{22}$ 
between atoms in the same component and $a_{12}$ between atoms 
in the different components as 
\begin{equation}
g_{jk} = \frac{2 \pi \hbar^{2} a_{jk}}{m_{jk}}
\end{equation}
with $m_{jk}^{-1} = m_{j}^{-1} + m_{k}^{-1}$. The GP model is given by the mean-field approximation 
for the many-body wave function and provides quantitatively good description of the 
static and dynamic properties of the dilute-gas BECs \cite{Pethickbook}.

%%%%%%%%%%%%%%%%%%%%%%%%%%%%%%%%%%%%%%
\subsection{Mapping to the nonlinear sigma model}
To derive the generalized NL$\sigma$M for two-component BECs 
from the GP energy functional (\ref{GPene}), we assume $m_{1}=m_{2}=m$ and $V_{1}=V_{2}=V$. 
By substituting the pseudospin representation Eq.(\ref{peudsospinrep}) of ${\bf \Psi}$, 
we obtain \cite{Kasamatsu2} 
\begin{eqnarray}
E = \int  d^3 x  \biggl\{ \frac{\hbar^{2}}{2m} \biggl[ 
(\nabla \sqrt{\rho})^{2}+ \frac{\rho}{4} \sum_{\alpha=1}^{3} 
(\nabla s_{\alpha})^{2} \biggr] + V \rho \nonumber \\
+ \frac{m \rho}{2} |{\bf v}_{\rm eff} |^{2}  
+  c_{0} + c_{1} s_{3} + c_{2} s_{3}^{2} \biggr\}, \label{gNLSM}
\end{eqnarray}
where we have introduced the effective superflow velocity
\begin{equation}
{\bf v}_{\rm eff} =\frac{ \hbar}{2m} ( \nabla \Theta - \cos \theta \nabla \phi ) 
\label{effvelociy}
\end{equation}
and the coefficients 
\begin{eqnarray}
c_{0} &=&\frac{ \rho }{8} [ \rho (g_{11}+g_{22}+2g_{12}) - 4 (\mu_{1} + \mu_{2}) ], \\
c_{1} &=&\frac{ \rho }{4}[  \rho (g_{11}-g_{22}) - 2 (\mu_{1} - \mu_{2}) ], \\
c_{2} &=&\frac{ \rho^{2}}{8} (g_{11}+g_{22}-2g_{12}). 
\end{eqnarray}
The coefficient $c_{1}$ can be interpreted as a longitudinal magnetic field 
that aligns the spin along the $x^3$-axis; it was assumed to be zero in this study. 
The term with the coefficient $c_{2}$ determines the spin--spin interaction 
associated with $s_{3}$; it is antiferromagnetic for $c_{2}>0$ and 
ferromagnetic for $c_{2}<0$ \cite{Kasamatsu2}.
Phase separation occurs for $c_{2}<0$, 
which we are focusing on. 
Further simplification can be achieved by assuming that 
$V=0$ and the total density is uniform 
through the relation $\rho = \mu / g$ where $g = g_{11} = g_{22}$ and 
$\mu = \mu_{1} = \mu_{2}$, and that the kinetic energy associated with the 
superflow ${\bf v}_{\rm eff} $ is negligible. 
Although the assumptions 
${\bf v}_{\rm eff}=0$ and $\rho$=const. become worse 
in the vicinities of vortex cores or domain walls, 
this simplification does not affect on later discussions 
about the vorton nucleations based on topology.

By using the healing length $\xi = \hbar/\sqrt{2mg \rho}$ 
as the length scale, the total energy can reduce to 
\begin{eqnarray}
&& \tilde{E} = \frac{E}{g \rho \xi^{3}} = \int  d^3 x \frac{1}{4} \left[ \sum_{\alpha=1}^3 (\nabla s_{\alpha})^{2} 
+ M^{2} (1 - s_{3}^{2}) \right] \label{nonsigmamod2}, \quad \quad \\
&& M^{2} = \frac{4 |c_{2}|}{g \rho^{2}},
\end{eqnarray}
where $M$ is the effective mass for $s_{3}$. This is a well-known massive 
NL$\sigma$M for effective description of a Heisenberg ferromagnet with spin--orbit coupling. 

Introducing a stereographic coordinate 
\begin{eqnarray}
 u={s_{1}-is_{2} \over 1-s_{3}},
\end{eqnarray} 
we can rewrite Eq. (\ref{nonsigmamod2}) as
\begin{equation}
\tilde{E} = \int d^3 x \frac{\sum_{\alpha=1}^{3} |\partial_{\alpha} u|^{2} +M^{2} |u|^{2}}
{(1+|u|^{2})^{2}}.
\label{nonsigmamod3}
\end{equation} 
Here, $u=0$ $(\infty)$ corresponds to the south (north) pole of the 
$S^2$ target space. 

%%%%%%%%%%%%%%%%%%%%%%%%%%%%%%%%%%%
\section{Wall -- anti-wall annihilation \label{sec:wall-anti-wall}} 
\subsection{Domain walls}
\begin{figure}[h]
\begin{center}
\includegraphics[width=0.5\linewidth,keepaspectratio]{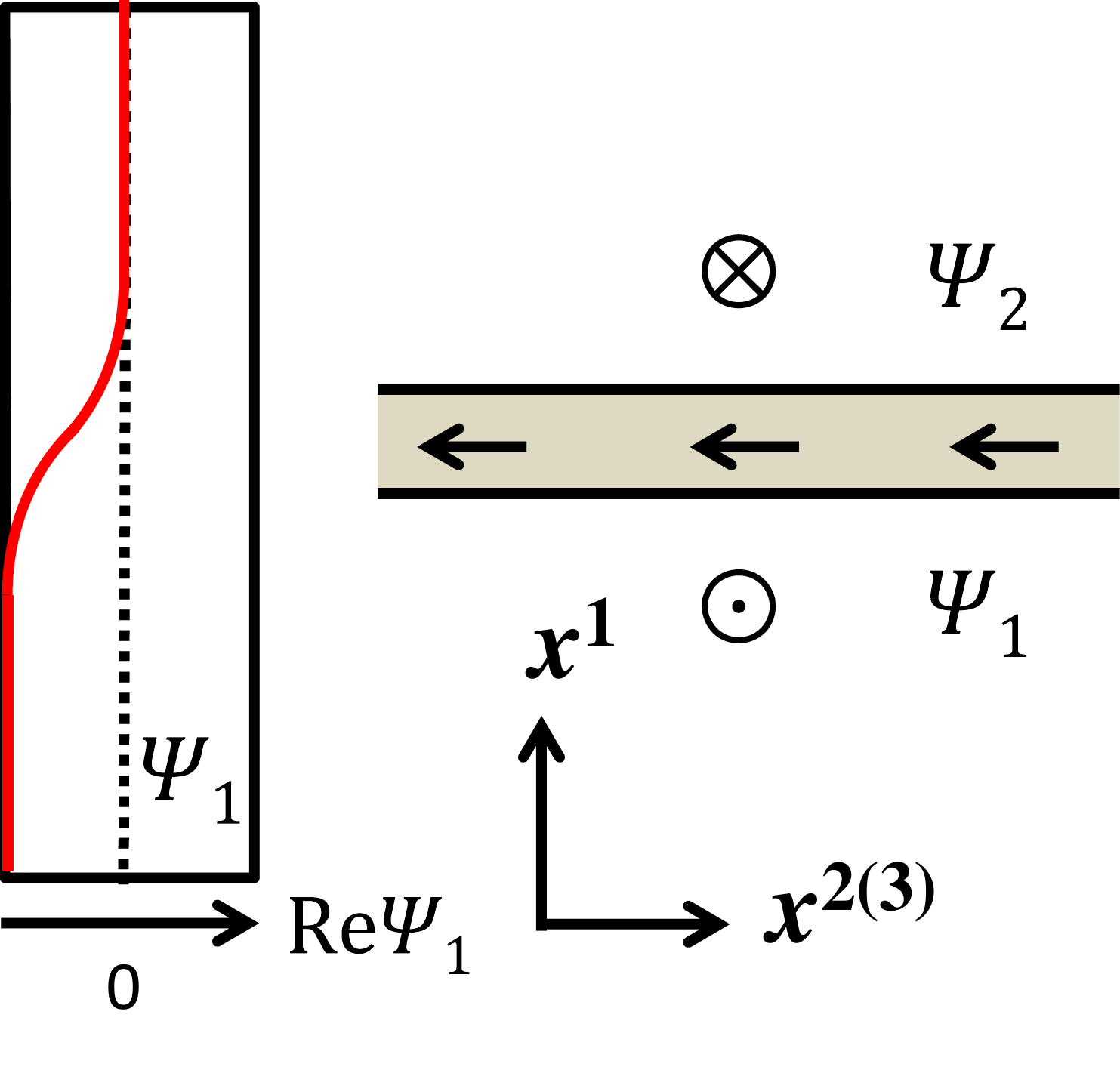}
\\

\hspace{-1.5cm} (a)\hspace{2cm}(b)

\caption{(Color online) A single domain wall 
in two-component BECs. 
(a) The amplitude of $\Psi_1$ for a domain wall.
(b) The pseudospin texture of 
the single domain wall perpendicular to the $x^1$-axis 
in real space.
The arrows denote points in the target space $S^2$. 
The gradient and interaction energies are 
localized around the wall, which is shaded schematically.
The arrows on the wall imply the phase $\phi_0$
which the wall possesses.
}
\label{fig:brane-2d} 
\end{center}
\end{figure}

\begin{figure}[h]
\begin{center}

\includegraphics[width=0.4\linewidth,keepaspectratio]{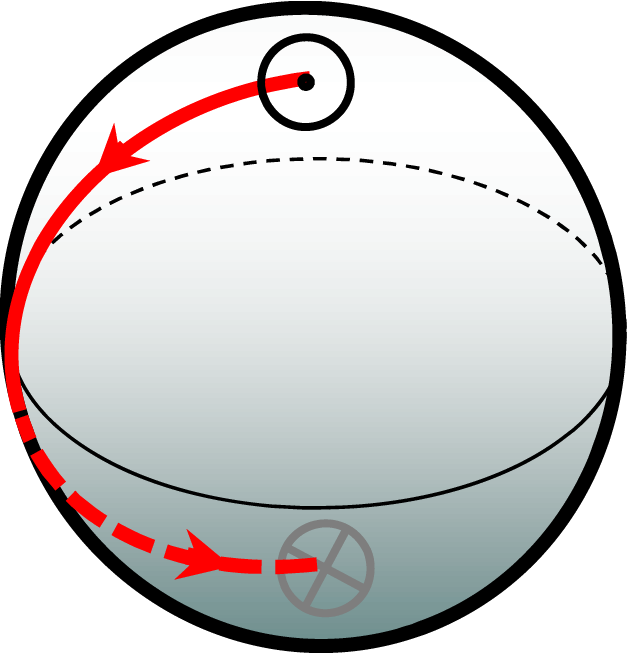}
\quad\quad
\includegraphics[width=0.4\linewidth,keepaspectratio]{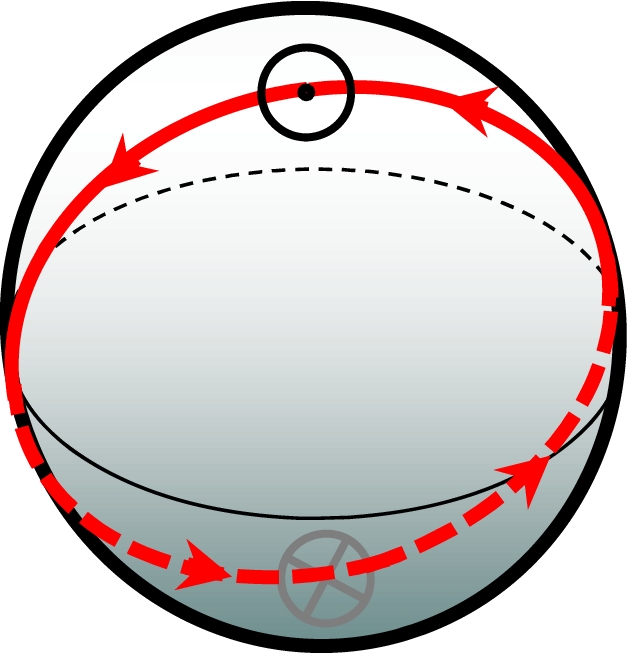}
\\
\hspace{-0.8cm} (a)\hspace{4cm}(b)

\caption{(Color online) The $S^2$ target space where 
the north and south poles are denoted by 
$\odot$ and $\otimes$, respectively.\\
(a) 
The path connecting the north and south poles 
represents the map from 
the path in the domain wall 
in Fig.~\ref{fig:brane-2d}(b) along the $x^1$-axis 
in real space from $x^1 \to - \infty$ to $x^1 \to + \infty$. 
The path in the $S^2$ target space 
passes through one point on the equator, 
which is represented by ``$\leftarrow$" in Fig.~\ref{fig:brane-2d}(b) 
in this example. 
In general, the $U(1)$ zero mode is localized on the wall. \\
(b) The path in the target space $S^2$ for 
a domain wall and an anti-domain wall. 
The path represents the map from 
the path along the $x^1$-axis from 
$x^1 \to - \infty$ to $x^1 \to + \infty$ in real space 
in Fig.~\ref{fig:brane-anti-brane-2d}(b).
}
\label{fig:brane-2d+} 
\end{center}
\end{figure}

For a domain wall perpendicular to the $x^1$-axis, $u=u(x^1)$,  
the total energy is bounded from below by the 
Bogomol'nyi--Prasad--Sommerfield (BPS) bound as 
\cite{Gauntlett:2000de,Isozumi:2004vg,Eto:2006pg,Kasamatsu:2010aq}
\begin{eqnarray}
\tilde{E} &=& \int d^3 x 
\frac{
|\partial_{1} u \mp Mu|^2 
\pm M (u^{\ast} \partial_1 u + u \partial_1 u^{\ast})}
{(1+|u|^{2})^{2}} \nonumber \\ 
 &\geq& |T_{\rm w}| \label{eq:BPS-bound-wall}
\end{eqnarray}
by the topological charge that characterizes the wall:
\begin{eqnarray}
T_{\rm w} = M \int  d^3 x \frac{ u^{\ast} \partial_1 u 
+ u \partial_1 u^{\ast}}{(1+|u|^{2})^{2}}   , 
\end{eqnarray}
where $\partial_i$ denotes 
the differentiation with respect to $x^i$. 
Among all configurations with a fixed boundary condition, 
{\it i.e.}, with a fixed topological charge $T_{\rm w}$, 
the most stable configurations with 
the least energy saturate the inequality (\ref{eq:BPS-bound-wall}) 
and satisfy the BPS equation  
\begin{equation}
\partial_{1} u \mp Mu =0,  
\end{equation}
which is obtained by $|...|^2=0$ in  Eq.~(\ref{eq:BPS-bound-wall}). 
This equation immediately gives the analytic form of the wall configuration
\begin{eqnarray}
u_{\rm w}(x^1)  = e^{\mp M (x^1-x^1_{0}) - i \phi_{0}}. \hspace{3mm}
\end{eqnarray}
The function $u_{\rm w}$ represents the domain wall with wall position 
$x^1_{0}$ and phase $\phi_{0}$ 
associated with $(s_{1}, s_{2})$; this phase $\phi_{0}$ yields 
the Nambu--Goldstone mode localized on the wall, 
as in Fig.~\ref{fig:brane-2d}. 
The sign $\mp$ implies a domain wall and an anti-domain wall.
The domain wall can be mapped to a path 
in the target space as shown in Fig.~\ref{fig:brane-2d+}(a).
%%%%%%%%%%%%%%%%%%%%%
%%%%%%%%%%%%%%%%%%%%%%%%%

%%%%%%%%%%%%%%%%%%%%%
%%%%%%%%%%%%%%%%%%%%%%%%%

\subsection{Wall anti-wall annihilation}
\begin{figure}[h]
\begin{center}

\includegraphics[width=0.5\linewidth,keepaspectratio]{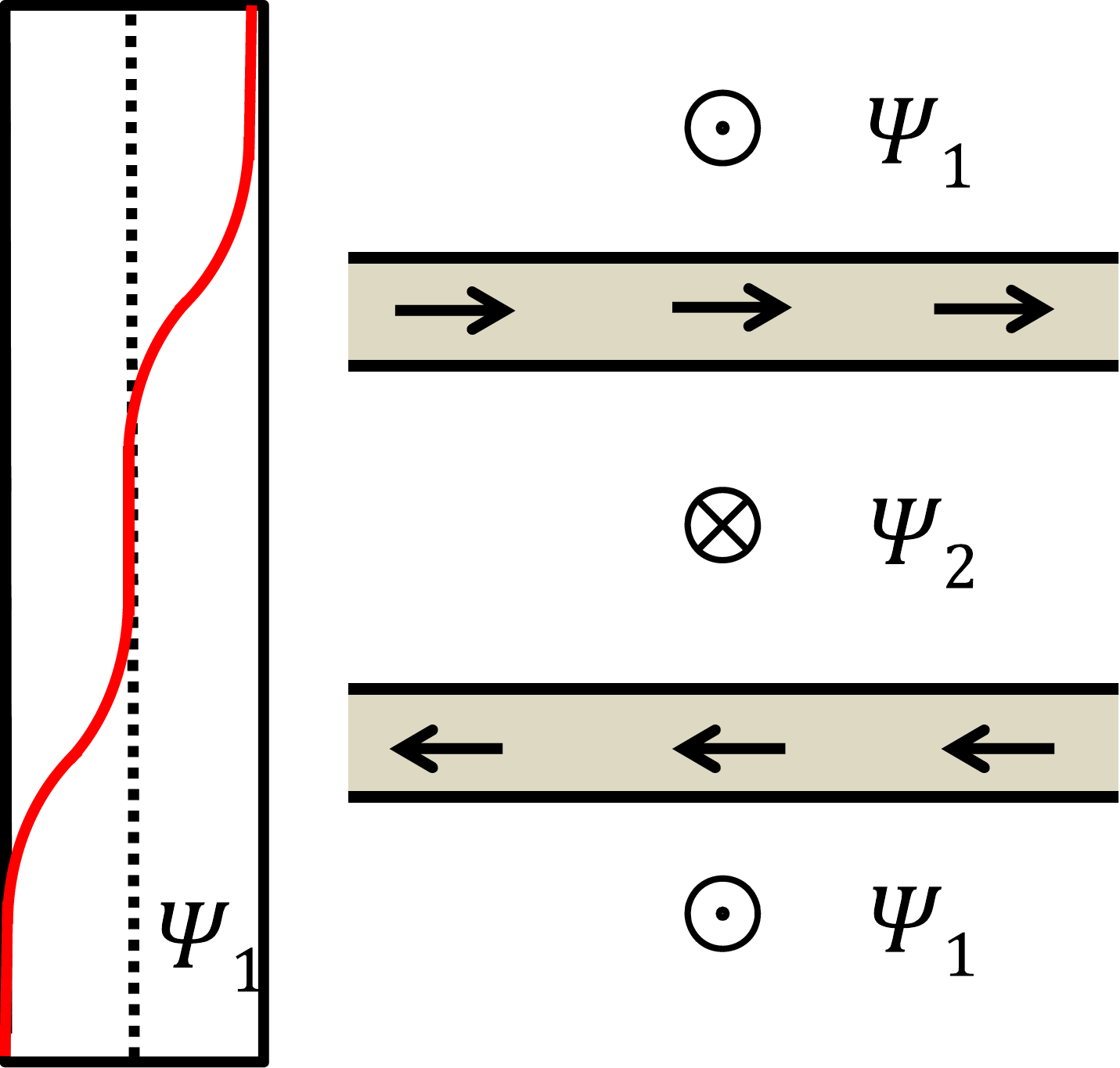}
\\
\hspace{-1.5cm} (a)\hspace{2cm}(b)

\caption{(Color online) A pair of domain wall and 
anti-domain wall in two-component BECs. 
(a) The amplitude of $\Psi_1$ for 
the wall and anti-wall configuration.
(b) The pseudo-spin texture of the wall and anti-wall 
configuration in real space.
The arrows denote the pseudo-spin. 
$\Psi_1$ ($\Psi_2$) is filled outside (between) the walls, 
where the other component is zero. 
In the upper (lower) region outside the walls, 
the phase of $\Psi_1$ is fixed to be zero ($\pi$). 
}
\label{fig:brane-anti-brane-2d} 
\end{center}
\end{figure}

As described in Sec.\ref{Intro}, 
we note that the intriguing experiment 
that mimicked the brane--antibrane annihilation was 
performed by Anderson {\it et al}. \cite{Anderson}.
They created the configuration shown in Fig.~\ref{fig:brane-anti-brane-2d}, 
where the nodal plane of a dark soliton in one component was filled 
with the other component. By selectively removing the filling component 
with a resonant laser beam, they made a planer dark-soliton in a 
single-component BEC. 
The dark soliton corresponds to the coincident limit 
of the two kinks in Fig.~\ref{fig:brane-anti-brane-2d}(a).  
It is known that the planer dark soliton in the 3D system 
is dynamically unstable for its transverse deformation 
(known as snake instability) \cite{Anderson}, which results in the decay of the dark soliton into vortex rings. 
%%%%%%%%%%%%%%%%%%%%%
%%%%%%%%%%%%%%%%%%%%%%

In our context, this experiment demonstrated the wall--anti-wall collision
and subsequent creation of cosmic strings, where the snake instability 
may correspond to ``tachyon condensation" in string theory \cite{Sen}. 
The procedure that removes the filling component can decrease
the distance $R$ between two domain walls and cause their collision.
% [see Fig.\ref{fig:3}]. 
The tachyon condensation can leave lower dimensional topological defects 
after the annihilation of D-brane and anti-D-brane. 
In our case of the phase-separated two-component BECs, the annihilation
of the 2-dimensional defects (domain walls) leaves 1-dimensional defects
(quantized vortices).
%%%%%%%%%%%%%%%%%%%%%
\begin{figure*}
\begin{center}

\includegraphics[width=0.18\linewidth,keepaspectratio]{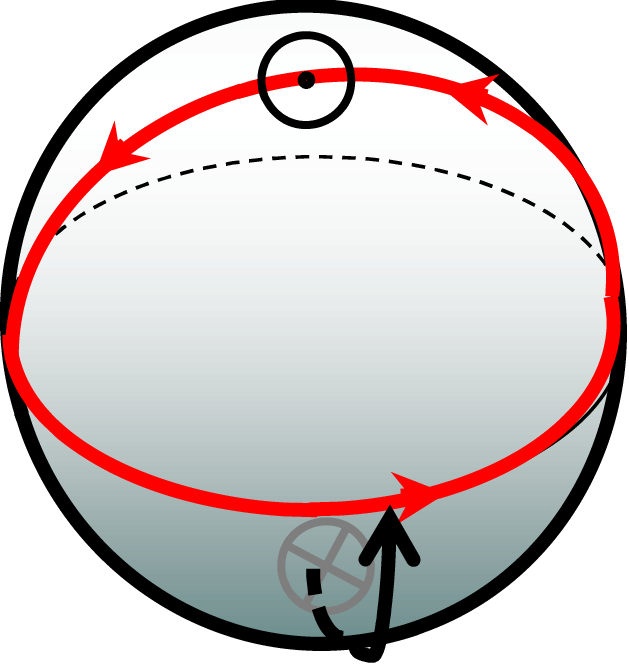}
\quad
\includegraphics[width=0.2\linewidth,keepaspectratio]{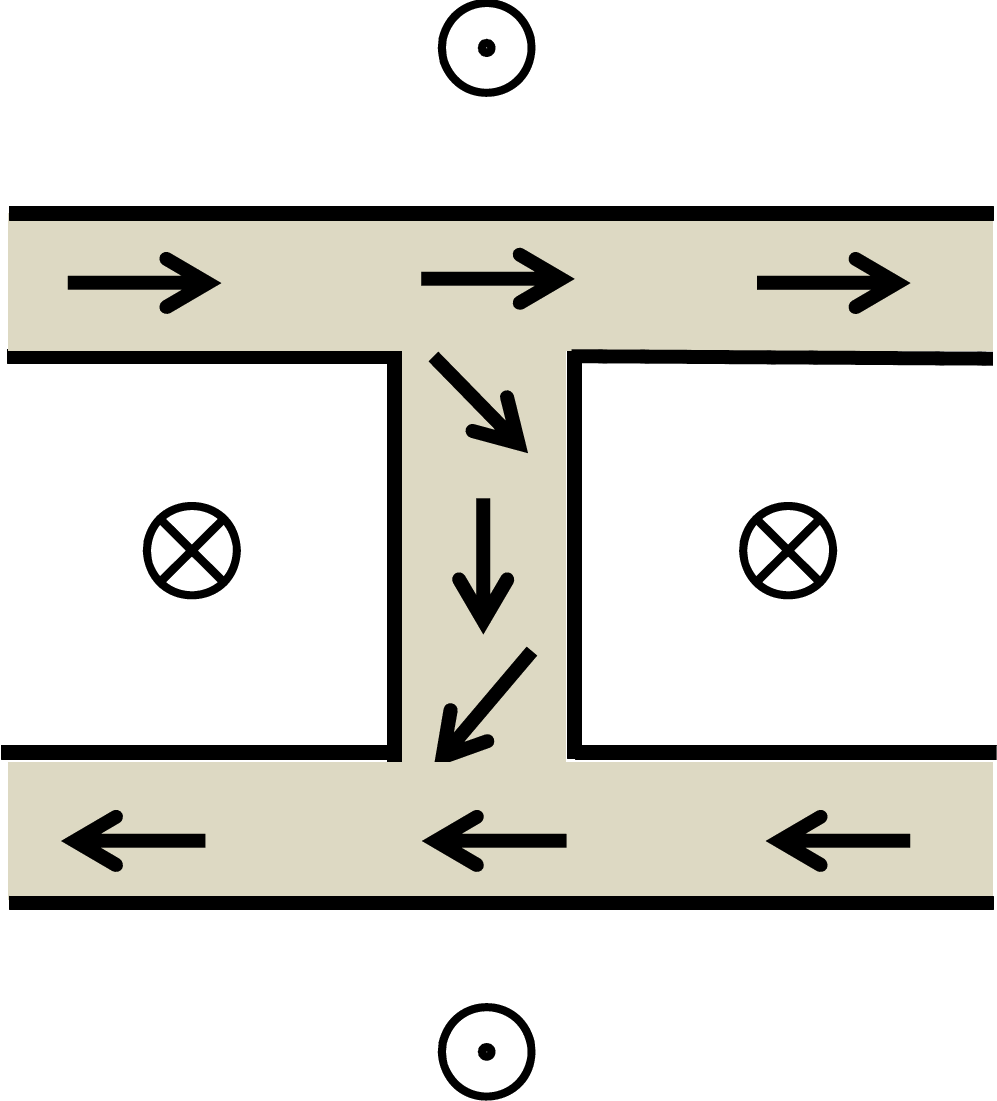}
\quad
\includegraphics[width=0.2\linewidth,keepaspectratio]{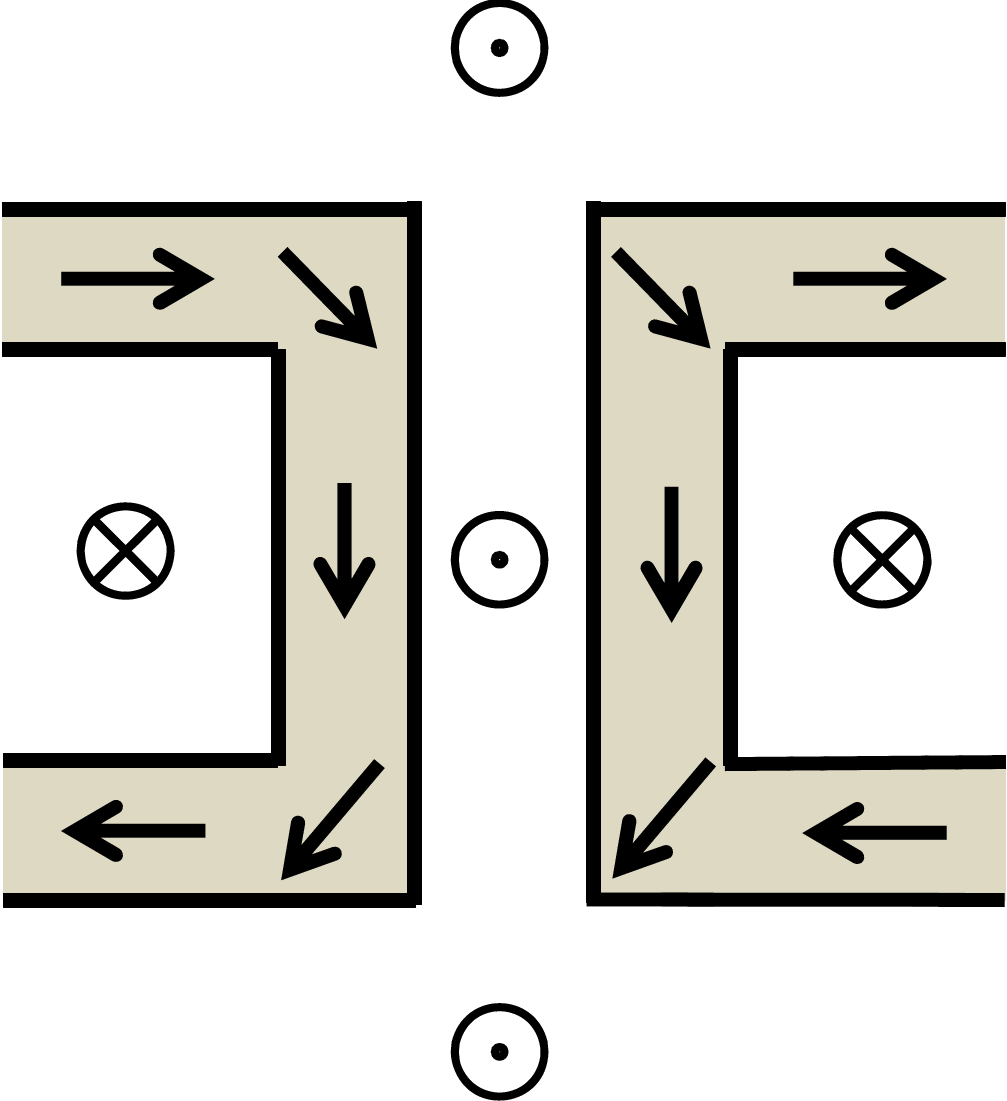}
\quad
\includegraphics[width=0.11\linewidth,keepaspectratio]{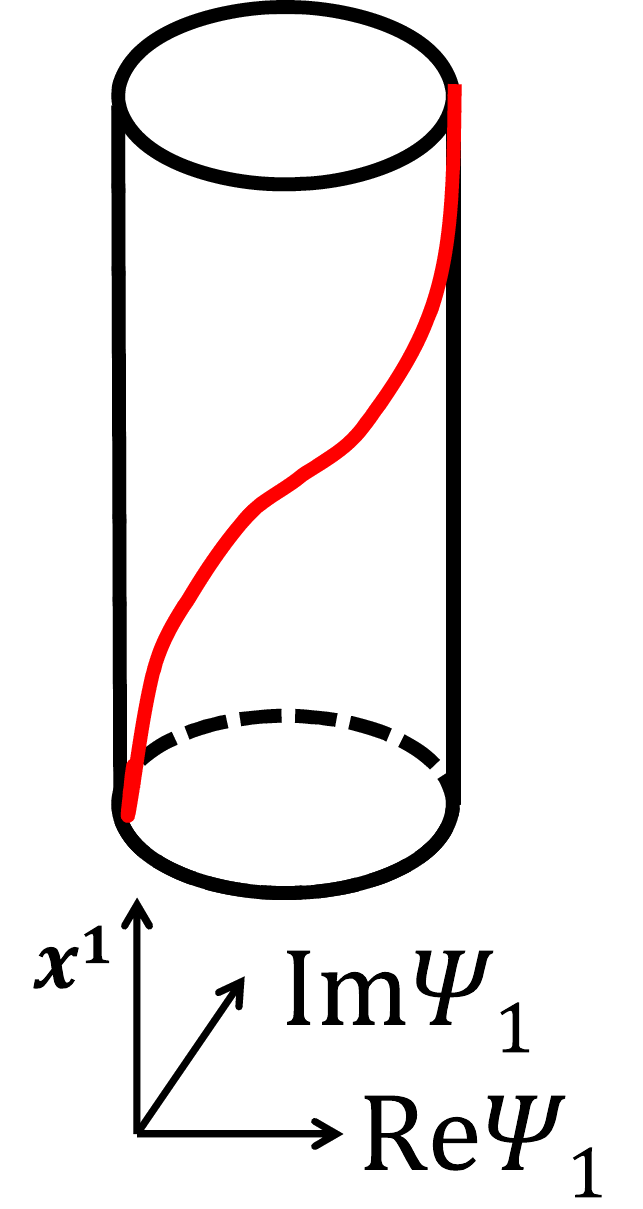}
\\
(a)\hspace{3.5cm} (c)\hspace{3.5cm} (e) \hspace{2.6cm}(g)\\

\medskip
\includegraphics[width=0.18\linewidth,keepaspectratio]{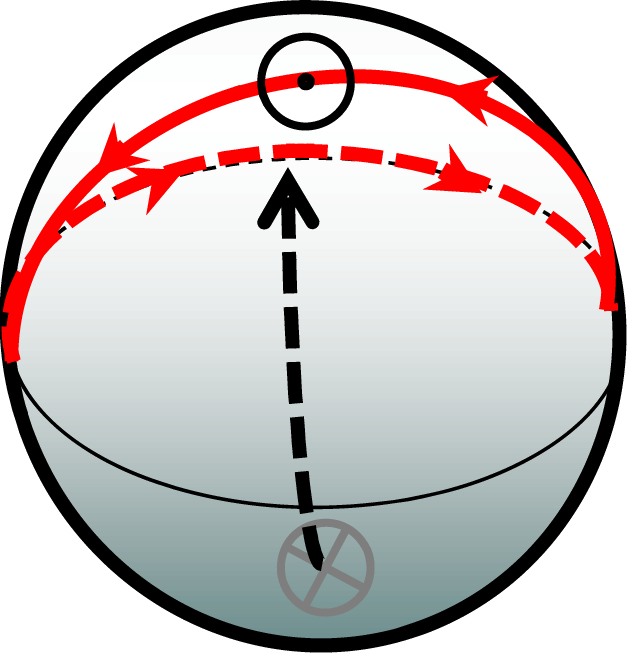}
\quad
\includegraphics[width=0.2\linewidth,keepaspectratio]{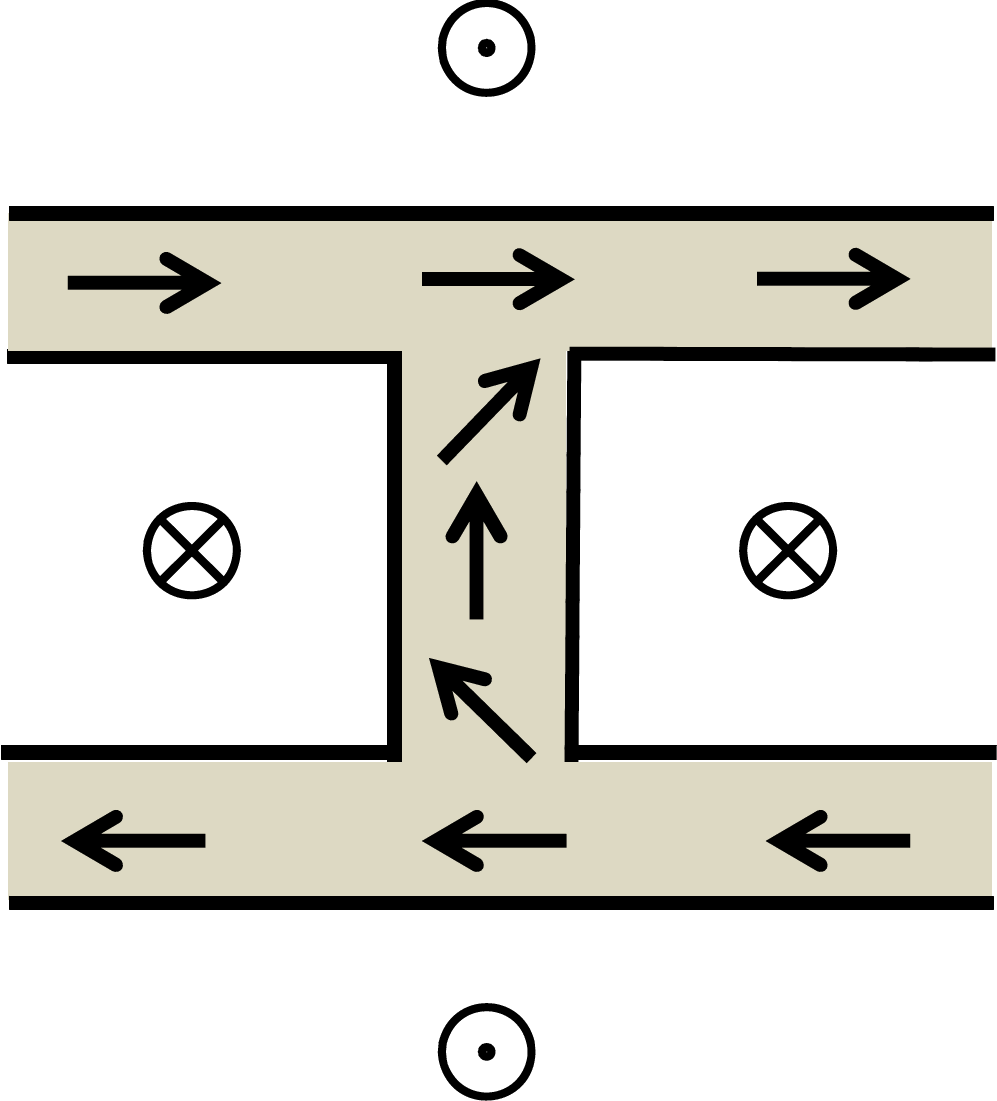}
\quad
\includegraphics[width=0.2\linewidth,keepaspectratio]{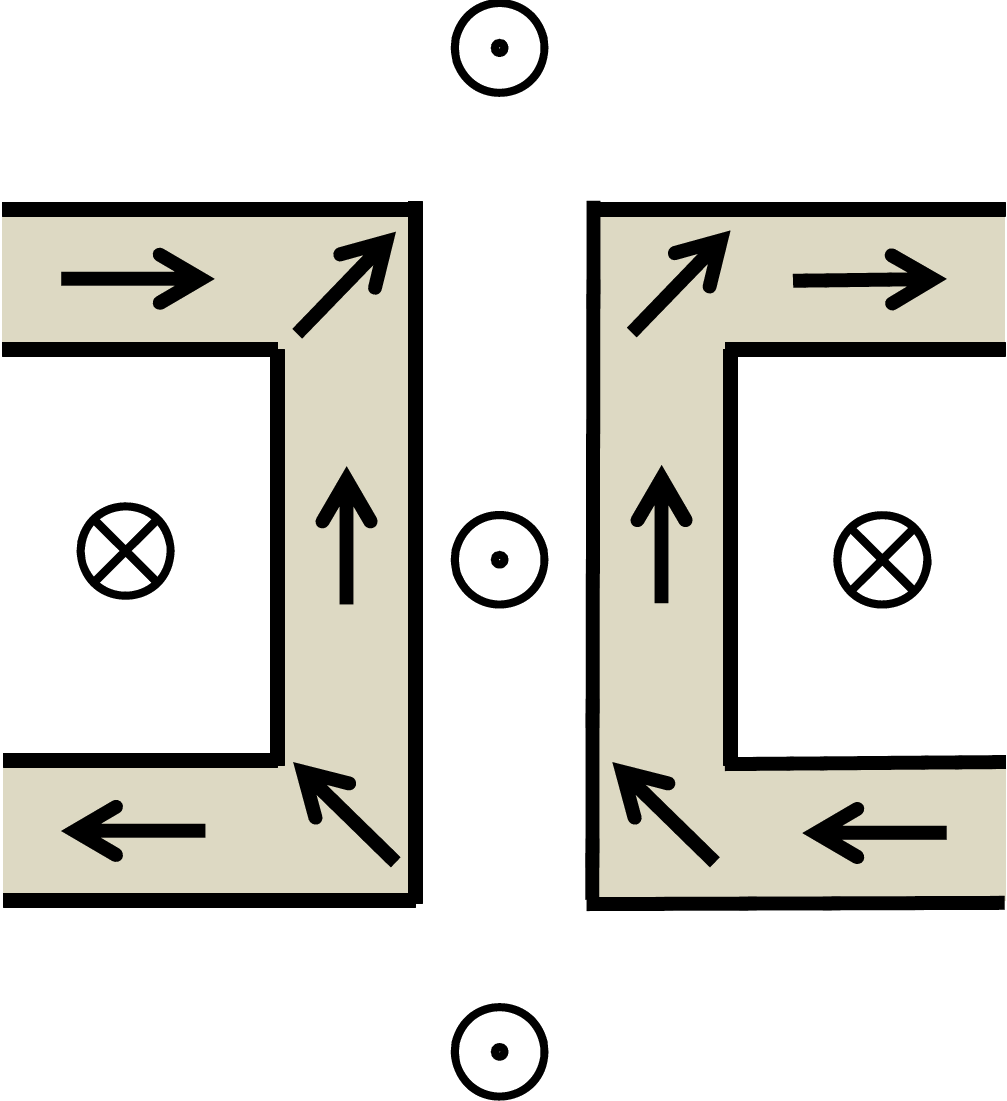}
\quad
\includegraphics[width=0.11\linewidth,keepaspectratio]{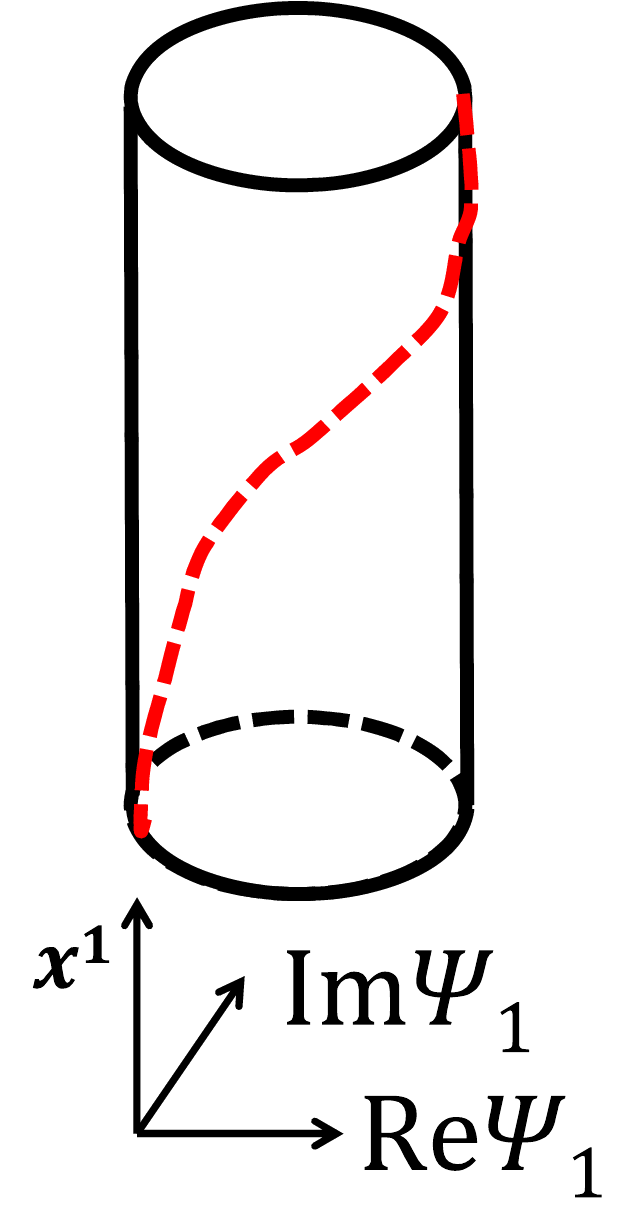}
\\
(b)\hspace{3.5cm} (d)\hspace{3.5cm} (f) \hspace{2.6cm}(h) 

\end{center}
\caption{(Color online)
Decaying processes of the wall-anti-wall pair.
(a,b) The loop in the pseudo-spin space 
is unwound in two ways.
(c,d) A bridge is created between the wall and anti-wall. 
In this process there are two possibilities of spin structure 
along the bridge. 
(e,f) The upper and lower regions are connected 
with a `passage' through the bridge being formed.
(g,h) The $\Psi_1$ component is filled inside the passage, 
and the phase of $\Psi_1$ component is connected 
anti-clockwise or clockwise.
} 
\label{fig:brane-anti-brane-anihilation} 
\end{figure*}
%%%%%%%%%%%%%%%%%%%%%

Let us discuss this in two dimensions in more detail.
Here $U(1)$ zero modes of the wall and the 
anti-wall are taken to be opposite 
as in Fig.~\ref{fig:brane-anti-brane-2d}~(b).
The configuration is mapped to a loop 
in the $S^2$ target space, 
see Fig.~\ref{fig:brane-2d+}~(b).
This configuration is unstable. 
It should end up with the vacuum with the up-spin $\odot$.
In the decaying process the loop is unwound from 
the south pole in the target space. 
To do this there are two topologically inequivalent ways, 
which are schematically shown in (a) and (b) 
in Fig.~\ref{fig:brane-anti-brane-anihilation}. 
In real space, 
at first, a bridge connecting two walls is created 
as in (c) and (d) in Fig.~\ref{fig:brane-anti-brane-anihilation}.
Here, there exist two possibilities of the spin structure of the bridge, 
corresponding to two ways of the unwinding processes. 
Along the bridge in the $x^1$-direction, 
the spin rotates (c) anti-clockwise or (d) clockwise 
on the equator of the $S^2$ target space. 
Let us label these two kinds of bridges by 
``$\downarrow$" and ``$\uparrow$", respectively.

In the next step, 
a `passage' through the bridge is formed 
as in (e) and (f) in Fig.~\ref{fig:brane-anti-brane-anihilation}, 
where the ground state, {\it i.e.}, the up-spin $\odot$ state, 
is filled between them. 
The phase of the filling $\Psi_1$ component through the passage 
is connected anti-clockwise or clockwise 
[Fig. \ref{fig:brane-anti-brane-anihilation} (g) and (h)] 
Let us again label these two kinds of passages by 
``$\downarrow$" and ``$\uparrow$", respectively. 
In either case, the two regions separated by the domain walls 
are connected through a passage created in the decay of domain walls.
Once created, these passages grow to holes in order 
to reduce the domain wall energy.

%%%%%%%%%%%%%%%%%%%%%
\begin{figure}[h]
\begin{center}
\includegraphics[width=0.4\linewidth,keepaspectratio]{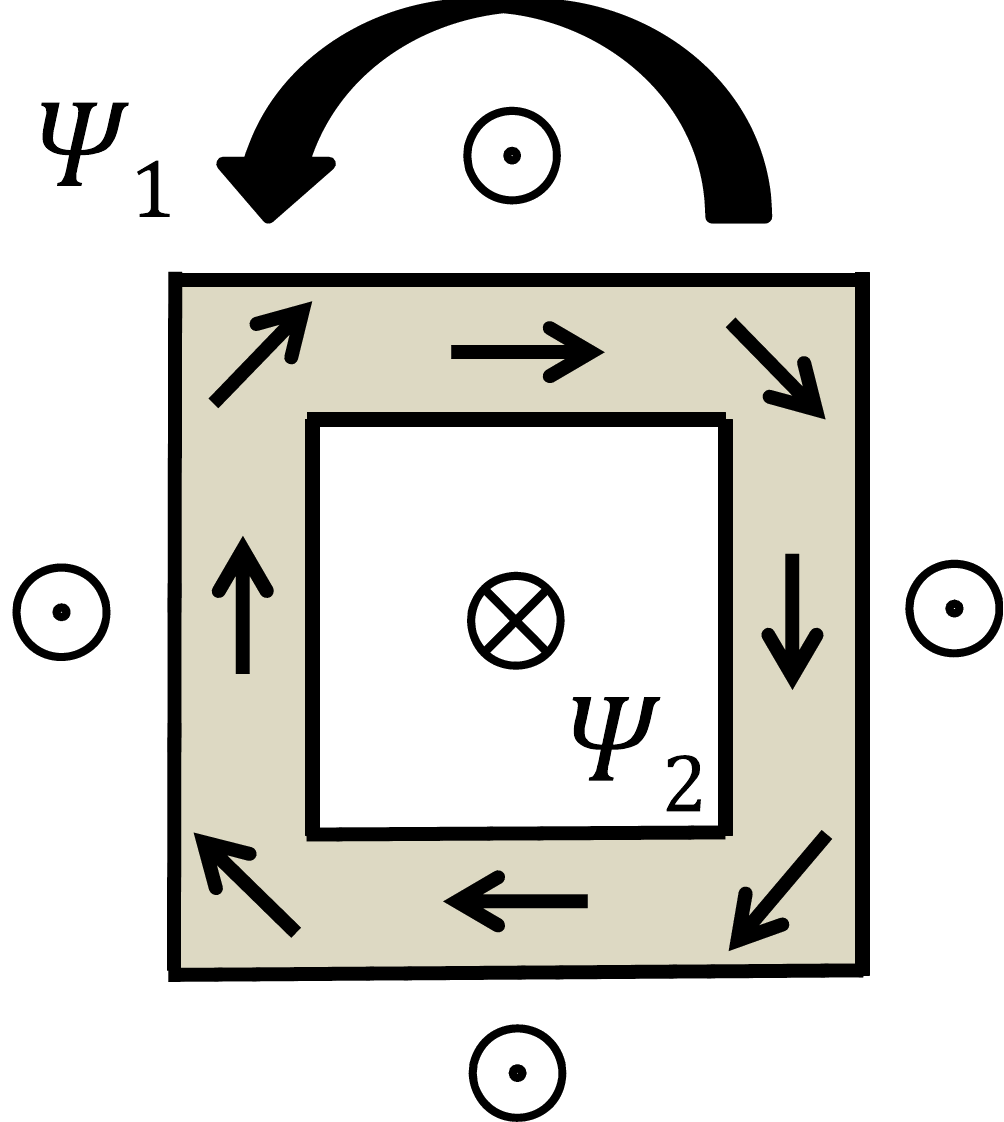}\quad
\includegraphics[width=0.4\linewidth,keepaspectratio]{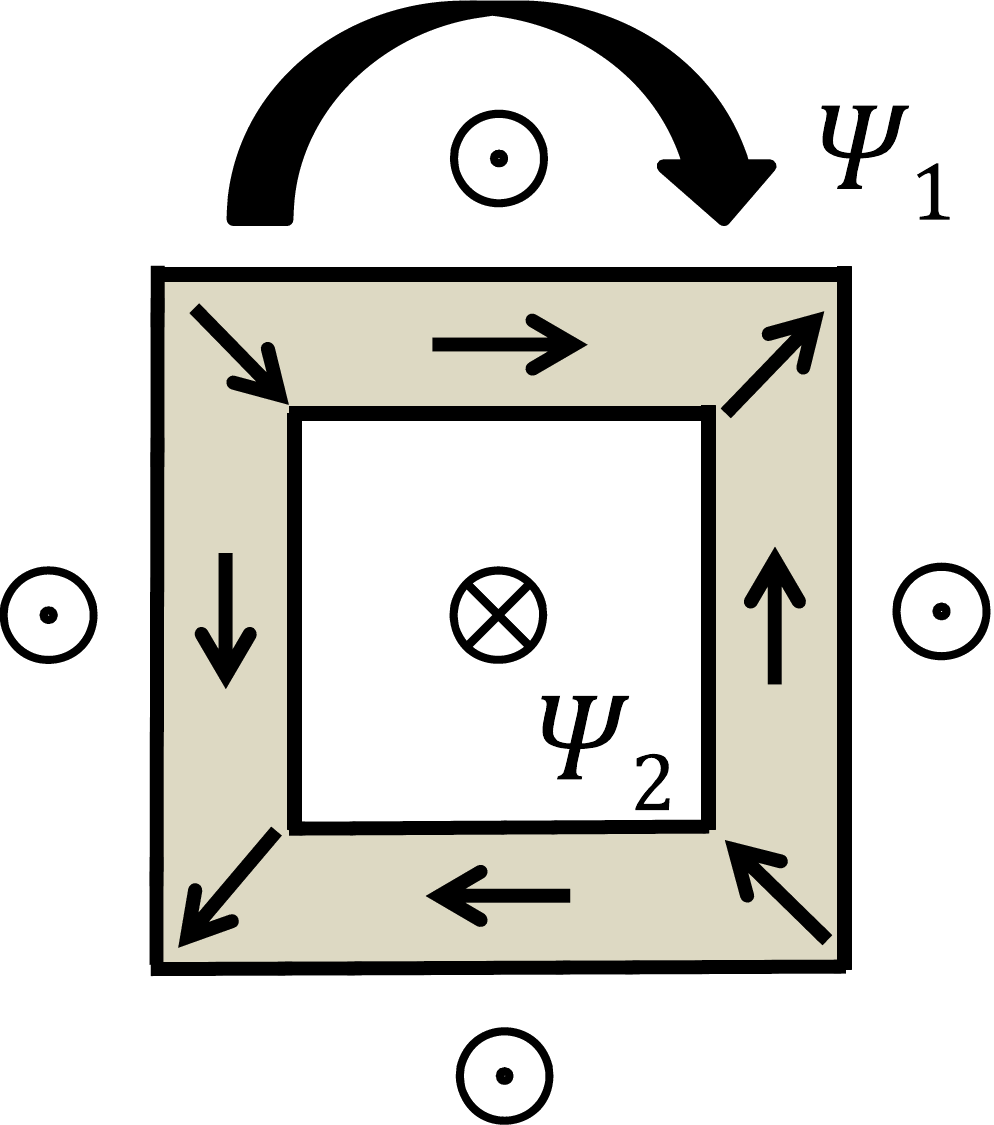}\\

\hspace{1.5cm}
(a)\hspace{3cm} (b)\hspace{1.5cm} 

\includegraphics[width=0.4\linewidth,keepaspectratio]{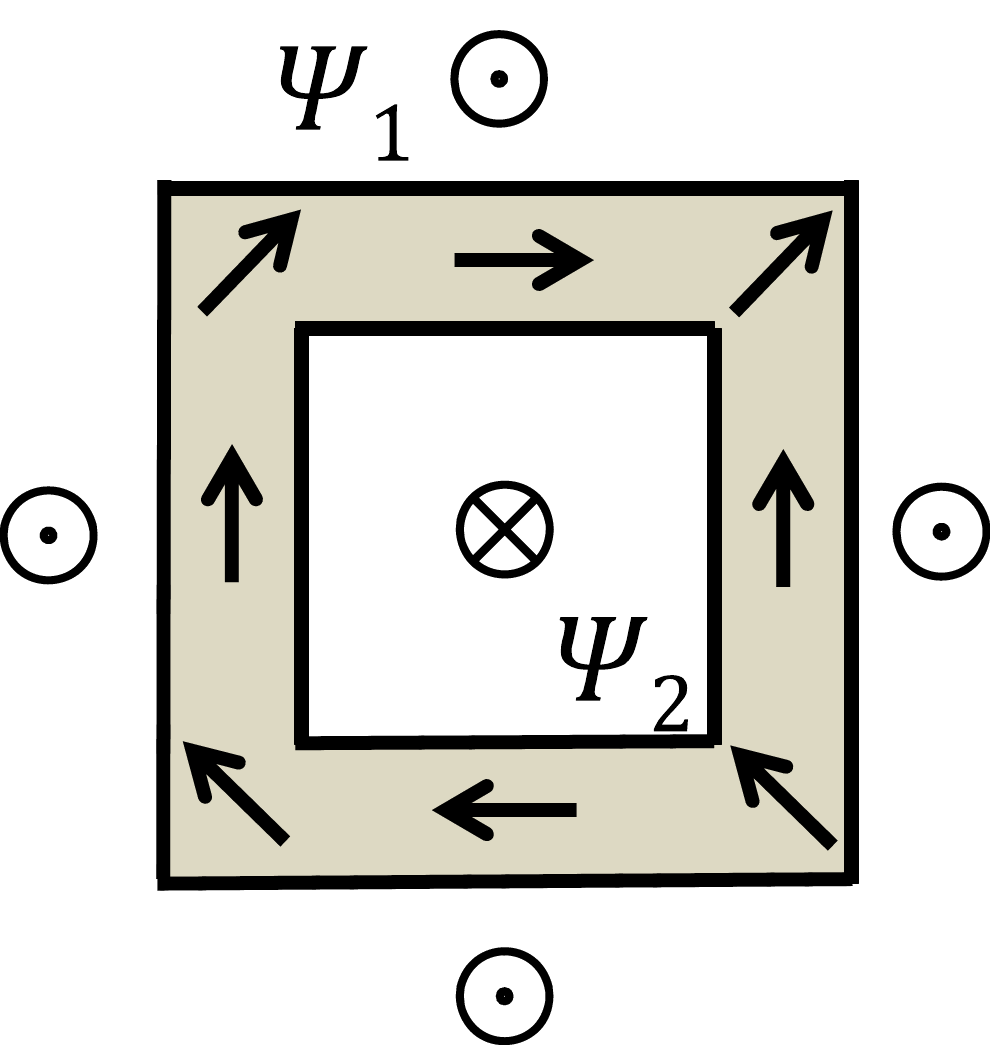}\quad
\includegraphics[width=0.4\linewidth,keepaspectratio]{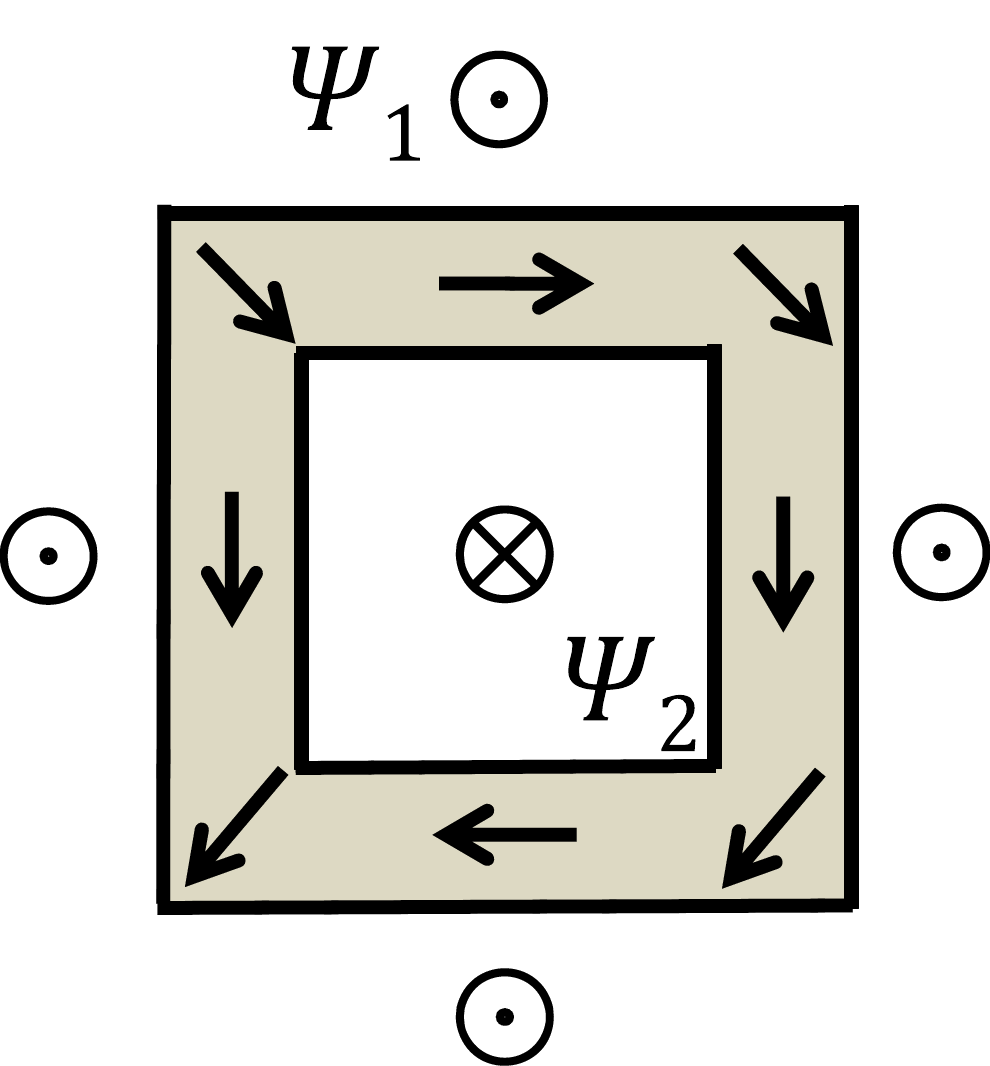}\\

\hspace{1.5cm}
(c)\hspace{3cm} (d)\hspace{1.5cm} 

\end{center}
\caption{(Color online) Stable and unstable wall rings. 
(a,b) Stable wall rings. The domain wall phase winds once 
(the winding number is $\pm 1$) along the rings. 
Total configurations are 2d skyrmions with 
a non-trivial element $\pm 1$ of the second homotopy group $\pi_2$.
The phase $\Psi_1$ winds $\pm 1$ so that it is a vortex, 
while $\Psi_2$ is filled inside the core. 
(c,d) Unstable wall rings. 
The domain wall phase does not wind  
(the winding number is $0$) along the rings. 
The phase $\Psi_1$ does not wind so that it cannot be a vortex, 
while $\Psi_2$ is filled inside the ring. 
They decay into ground state (up pseudo-spin).
} 
\label{fig:winding} 
\end{figure}
%%%%%%%%%%%%%%%%%%%%%
Several holes are created in the entire decaying process.
Let us focus a pair of two neighboring holes. 
Then, one can find a ring of a domain wall between the holes 
as shown in Fig.~\ref{fig:winding}.
Here, since there exist two kinds of holes ($\uparrow$ and $\downarrow$), 
there exist four possibilities of the rings,  
(a) $\uparrow\downarrow$, (b) $\downarrow\uparrow$, 
(c) $\uparrow\uparrow$ and (d) $\downarrow\downarrow$  
in Fig.~\ref{fig:winding}.
In all the cases, the $\Psi_2$ component is confined 
in the domain wall rings.
The phase of $\Psi_1$ component has a nontrivial winding 
outside the rings of types (a) and (b),
whereas it does not have a winding outiside 
the domain wall rings of types (c) and (d). 
Consequently, the domain wall rings of types (c) and (d)
can decay and end up with the ground state $\odot$. 
However, the decay of the rings of types 
(a) and (b) is topologically forbidden; 
they are nothing but coreless vortices.

In the $O(3)$ NL$\sigma$M, 
the domain wall rings of types (a) of (b)
are the Anderson--Toulouse vortices \cite{AndersonToulouse}, 
or lumps in field theory \cite{Polyakov:1975yp}. 
The solutions can be written 
as ($z \equiv x^1+ix^2$)
\begin{equation}
 u = u_0 = \sum_{i=1}^k{\lambda_i \over z-z_i }\quad \mbox{or} 
\quad u = \bar u_0  \label{eq:lump}
\end{equation} 
for a lump or an anti-lump, 
where $z_i \in {\bf C}$ represent the position of the lump 
while  
and $\lambda_i \in {\bf C}^*$ with  
$|\lambda_i|$ and $\arg \lambda_i$
representing the size and the $U(1)$ orientation of 
the lump, respectively. 
In fact, one can show that 
these configurations have a nontrivial winding 
in the second homotopy group $\pi_2 (S^2)\simeq {\bf Z}$ 
which can be calculated from 
\begin{eqnarray}
{1\over 2\pi}\int d^2 x \,
  {i (\partial_1 u^* \partial_2 u 
    - \partial_2 u^* \partial_1 u )
     \over  (1+|u|^2)^2}.
\end{eqnarray}
The wall rings of
(a) and (b) in Fig.~\ref{fig:winding} 
belong to $+1$ and $-1$ of $\pi_2 (S^2)$, respectively. 
Namely they are a lump and an anti-lump, respectively.

\begin{figure}
\begin{center}
\includegraphics[width=1\linewidth,keepaspectratio]{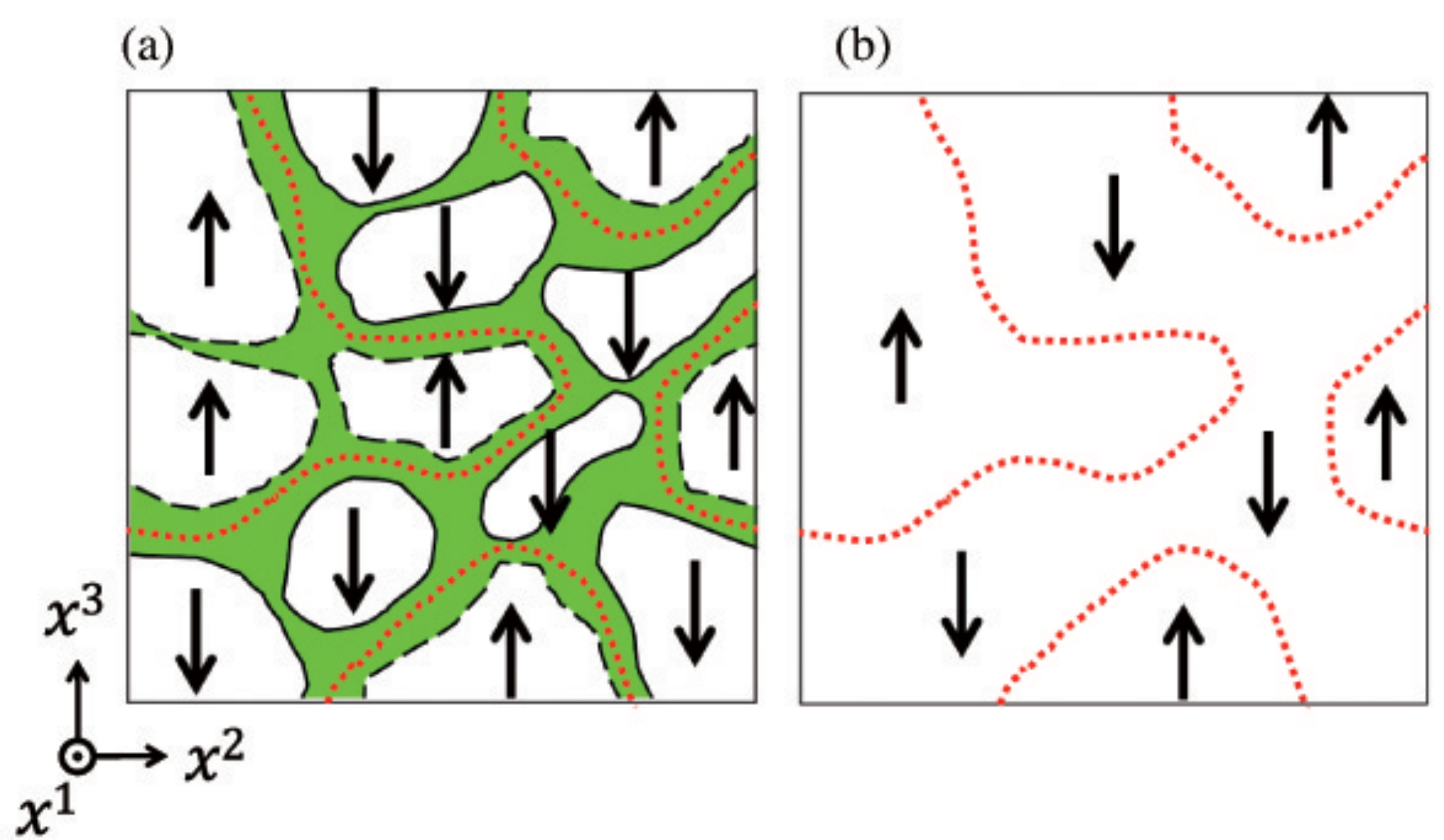}
\end{center}
\caption{(Color online) Decay of a domain wall pair in three dimensions. 
Domain walls are two-dimensional objects.
(a) Two kinds of holes labeled by $\uparrow$ and $\downarrow$ 
are created after a domain-wall pair decay.
(b) Along the boundary of these two kinds of holes, 
there appear vortex lines (denoted by dotted lines), 
which in general making vortex loops. 
\label{fig:2d-hole} 
} 
\end{figure}

So far we have discussed two dimensional space 
in which domain wall is a line and a vortex is point-like.
In three dimensions, 
domain walls have two spatial dimensions. 
When the decay of the domain wall pair occurs,
there appear two-dimensional 
holes, which can be labeled by 
$\downarrow$ or $\uparrow$ in 
Fig.~\ref{fig:2d-hole}(a).
Along the boundary of these two kinds of holes, 
there appear vortex lines, 
which in general making vortex loops,  
as in Fig.~\ref{fig:2d-hole}(b). 
This process has been numerically demonstrated 
\cite{Takeuchi:2011}.
The vortex rings decay into the fundamental excitations 
in the end.
%%%%%%%%%%%%%%%%%%%%%
%%%%%%%%%%%%%%%%%%%%%

%%%%%%%%%%%%%%%%%%%%%%%%%%%%%%%%%%%%%%%%%%%%%%%%%%%%%
\section{D-brane -- anti-D-brane annihilation  
\label{sec:brane-anti-brane} }

%%%%%%%%%%%%%%%
\subsection{D-brane soliton}
The D-brane soliton by Gauntlett {\it et. al}. \cite{Gauntlett:2000de} can be 
reproduced in two-component BECs as follows \cite{Kasamatsu:2010aq}. 
For a fixed topological sector, vortices (a domain wall) parallel (perpendicular) to the $x^1$-axis, 
the total energy is bounded from below by the BPS bound as 
\cite{Gauntlett:2000de,Isozumi:2004vg,Eto:2006pg,Kasamatsu:2010aq}
\begin{eqnarray}
\tilde{E}
&=& \int  d^3 x 
\frac{
 |\partial_{1} u \mp Mu|^2 
+|(\partial_{2} \mp i \partial_{3}) u|^2}{(1+|u|^{2})^{2}} \nonumber\\ 
&\pm& \int  d^3 x 
\frac{ M(u^{\ast} \partial_{1} u + u \partial_{1} u^{\ast})
      + i(\partial_{2} u^{\ast} \partial_{3} u 
       - \partial_{3} u^{\ast} \partial_{2} u)}
     {(1+|u|^{2})^{2}} 
   \nonumber\\ 
 &\geq& |T_{\rm w}| + |T_{\rm v}|
\end{eqnarray}
 by the topological charges 
that characterize the wall and vortices:
\begin{eqnarray}
T_{\rm w} &=& M \int  d^3 x \frac{ u^{\ast} \partial_{1} u + u \partial_{1} u^{\ast}}{(1+|u|^{2})^{2}}   ,  \\
T_{\rm v} &=& i \int  d^3 x  \frac{\partial_{2} u^{\ast} \partial_{3} u - \partial_{3} u^{\ast} \partial_{2} u }{(1+|u|^{2})^{2}} . 
\end{eqnarray}
Then, the least energy configurations with 
fixed topological charges (a wall with a fixed number of vortices) 
satisfy the BPS equations 
\begin{equation}
\partial_{1} u \mp Mu =0, \hspace{6mm} 
(\partial_{2} \mp i \partial_{3}) u = 0 .
\end{equation}
The analytic form of the wall--vortex composite solitons 
can be found 
($z\equiv x^2+ix^3$)
\begin{equation}
 u(x^1,z)=u_{\rm w}(x^1) u_{\rm v}(z), 
 \label{wallvortexcom}
\end{equation} 
where \cite{Isozumi:2004vg}
\begin{eqnarray}
u_{\rm w}(x^1)  = e^{\mp M (x^1-x^1_{0}) - i \phi_{0}},
\\ \hspace{3mm}
u_{\rm v}(z) = 
\frac{\prod_{j=1}^{N_{v_1}} (z - z_{j}^{(1)})}
{\prod_{j=1}^{N_{v_2}} (z - z_{j}^{(2)})}.
\label{wallvortexcomp}
\end{eqnarray}
The function $u_{\rm w}$ represents the domain wall with 
wall position $x^1_{0}$ 
and phase $\phi_{0}$. 
The function $u_{\rm v}$ gives the vortex configuration, being written by arbitrary analytic functions 
of $z$; the numerator represents $N_{v_1}$ vortices 
in one domain ($\Psi_{1}$ component) and 
the denominator represents $N_{v_2}$ vortices in the 
other domain ($\Psi_{2}$ component). The positions of 
the vortices are denoted by $z_{j}^{(1)}$ and $z_{j}^{(2)}$.
The total energy does not depend on the form of the solution, 
but only on the topological charges 
as $T_{\rm w} = \pm M$ or 0 (per unit area), and $T_{\rm v} = 2 \pi N_{\rm v}$ 
(per unit length), where 
$N_{\rm v}$ is the number of vortices passing through a certain $x^1=$ const plane.

%%%%%%%%%%%%%%%%%%%%%%%%%
\begin{figure}[t]
\begin{center}
\includegraphics[width=1.0 \linewidth,keepaspectratio]{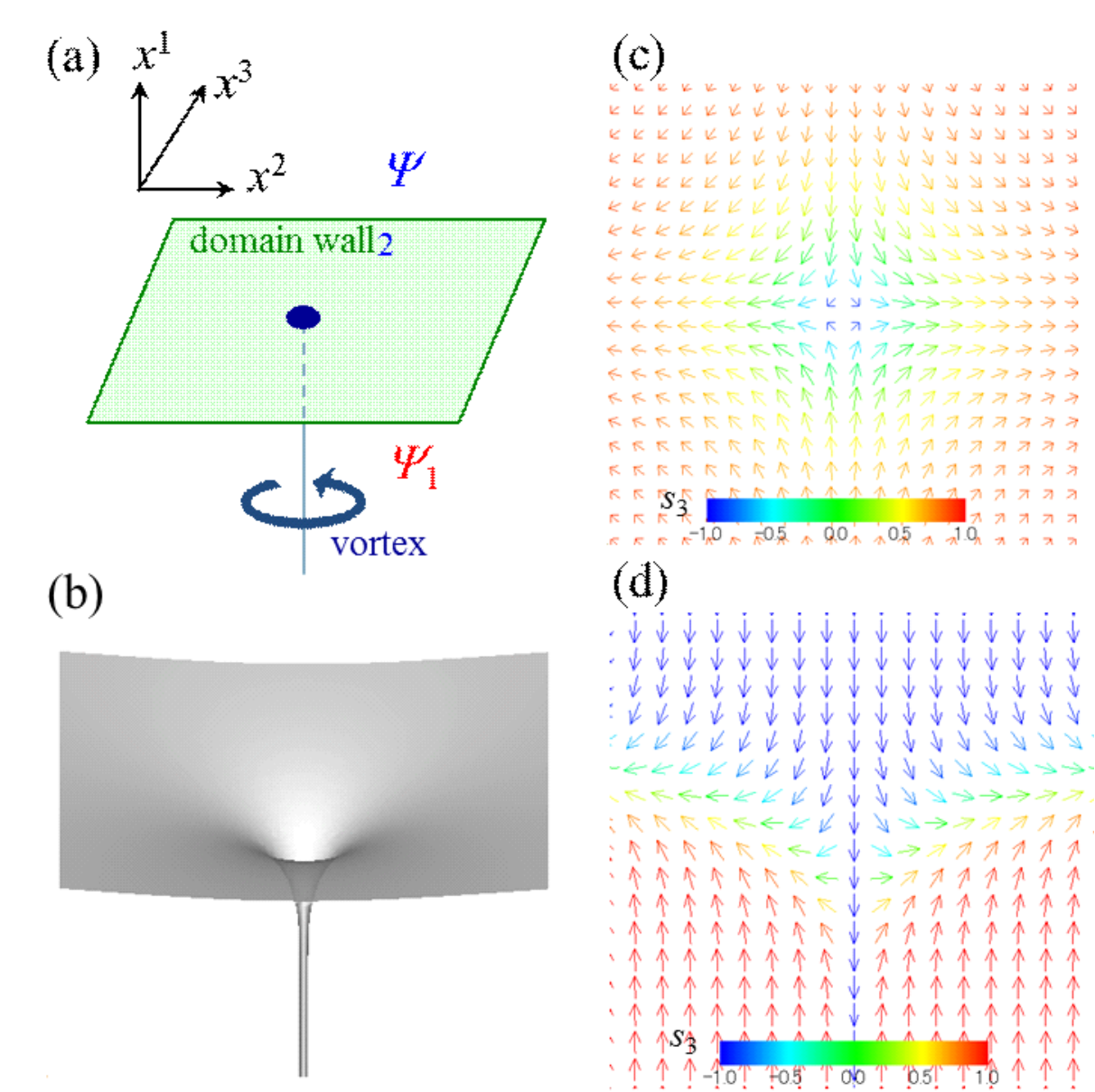}
\caption{(Color online) The typical D-brane soliton in two-component BECs. 
(a) Schematic illustration of the wall-vortex soliton 
configuration viewed from the length scale larger 
than the domain-wall width and the vortex core size. The two-component BECs 
$\Psi_{1}$ ($x^1<0$) and $\Psi_{2}$ ($x^1>0$) are separated by the domain 
wall. A single vortex located at $x^1<0$ ($\Psi_{1}$ component) 
is connected to the domain wall. 
(b) The isosurface of $s_3=0$ for the solution Eq. (\ref{wallvortexcom}) 
of the NL$\sigma$M, where $M=1$, 
$x_0^1=0$, $\phi_0=0$, $N_{v1}=1$, $N_{v2}=0$, and $z_1^{(1)} = 0$. 
The corresponding spin textures ${\bf s}$ in the $z=0$ plane 
and $y=0$ plane are shown in (c) and (d), respectively. 
The magnitude of $s_{3}$ is denoted by color. 
We have ${\bf s} = (0,0,-1)$ along the vortex core. 
\label{Dbrain1}} 
\end{center}
\end{figure}
\begin{figure}[t]
\begin{center}
\includegraphics[width=1.0 \linewidth,keepaspectratio]{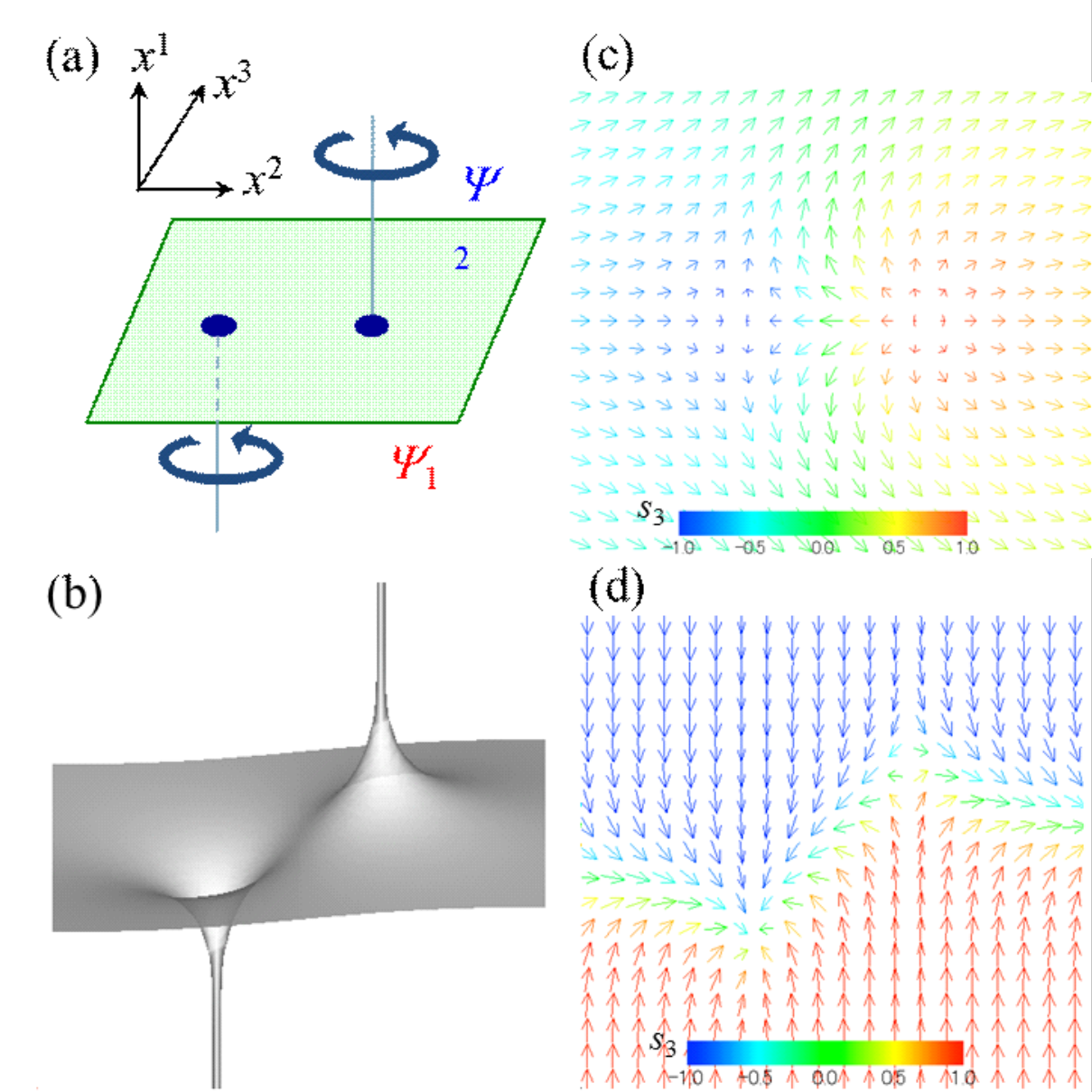}
\caption{(Color online) The D-brane soliton to which two vortices attach. 
(a) Schematic illustration of the configuration in which 
each component has a single vortex connected to the wall. 
(b) The isosurface of $s_3=0$ for the solution Eq. (\ref{wallvortexcom}) 
of the NL$\sigma$M, where $M=1$, 
$x_0^1=0$, $\phi_0=0$, $N_{v1}=1$, $N_{v2}=1$, $z_1^{(1)} = -2$, and $z_1^{(2)} = 2$. 
The corresponding spin textures ${\bf s}$ in the $z=0$ plane 
and $y=0$ plane are shown in (c) and (d), respectively. 
The magnitude of $s_{3}$ is denoted by color.  
The wall becomes asymptotically flat due to the balance 
between the tensions of the attached vortices.
\label{fig:2} } 
\end{center}
\end{figure}

%%%%%%%%%%%%%%%%%%%%%%%%%%%%%%%%%%
Figure \ref{Dbrain1} shows a D-brane soliton with the simplest wall--vortex 
configuration of Eq. (\ref{wallvortexcom}). 
A vortex exists in $x^1<0$ and forms a texture, 
where the spin points down at the center and rotates continuously from down to up as it moves radially outward. 
The edge of vortex attaches to the wall, causing it to bend 
logarithmically as $x^1=\log |z|/M$ [Fig.~\ref{Dbrain1}(b) and (d)]. 
We can construct solutions in which an arbitrary number of 
vortices are connected to the domain wall by multiplying by the additional factors
$z-z_{j}^{(i)}$ [see Eq. (\ref{wallvortexcomp})]; 
Fig. \ref{fig:2} shows a solution in 
which both components have one vortex connected to the wall. 
In the NL$\sigma$M, the energy is independent of the vortex positions 
$z_{j}^{(i)}$ 
on the domain wall; in other words, there is no static interaction between vortices. 
%%%%%%%%%%%%%%%%%%%%%%%%%%%%%%%%%%%%%%%%

%%%%%%%%%%%%%%%%%%%%%%%%%%%%%%%%%%%%%%%%%%%%%%%%%%%%%%%%%%%%%%
\subsection{Brane-anti-brane annihilation with a string}

%%%%%%%%%%%%%%%%%%%%%
\begin{figure*}
\begin{center}
\includegraphics[width=1\linewidth,keepaspectratio]{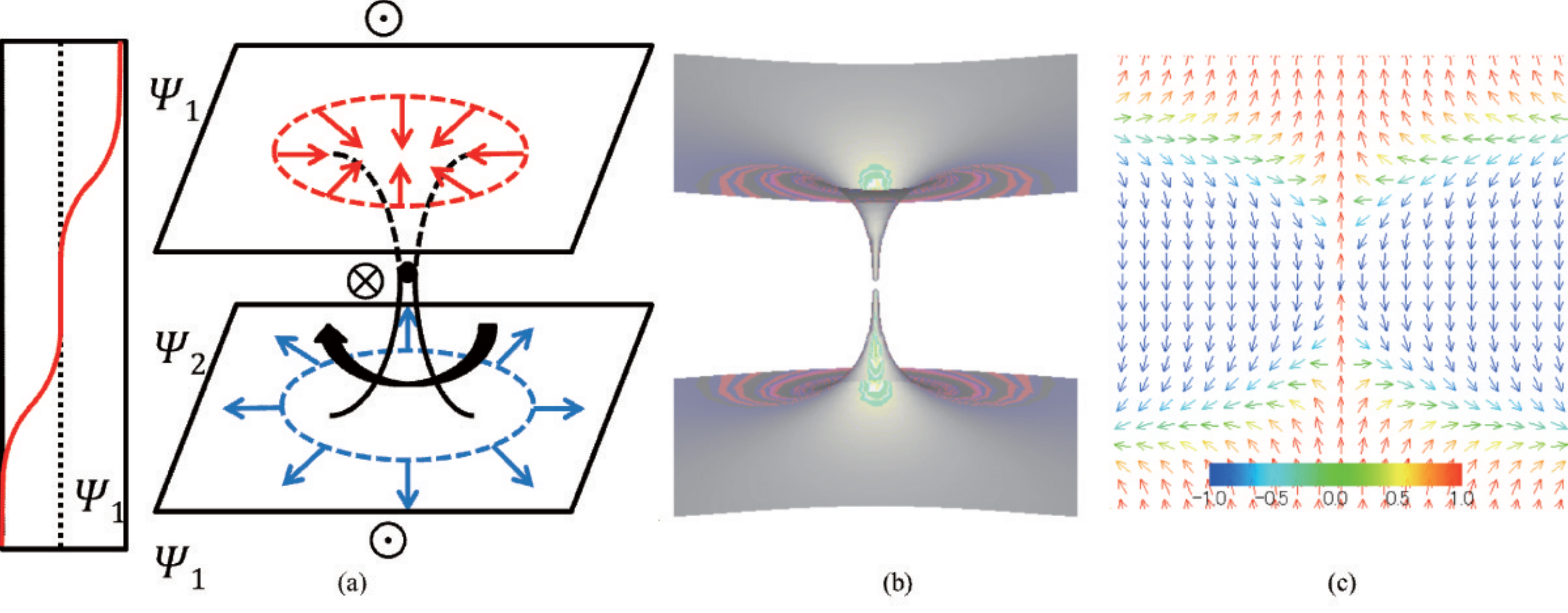}
\end{center}
\caption{(Color online) (a) A pair of a D-brane (domain wall) and 
an anti-D-brane (anti-domain wall) stretched by a 
string (vortex) in two-component BECs. 
The branes are perpendicular to the $x^1$-axis 
and the string is placed along the $x^1$-axis.
The arrows denote pseudo-spins. 
The $\Psi_1$ ($\Psi_2$) component is filled outside 
(between) the branes, 
where the other component is zero. 
In the upper (lower) region outside the branes, 
the phase of $\Psi_1$ is fixed to be zero ($\pi$). 
In the middle region, the phase of $\Psi_2$ has winding around 
the vortex placed at the $x^1$-axis. 
Accordingly, the pseudo-spin rotates once (anti-)clockwise 
at the endpoint of string on the (anti-)brane. 
The profile of the $\Psi_1$ component 
along a line parallel to the $x^1$-axis at $(x^2,x^3) \neq 0$
represents two kinks in the left pannel, 
while the profile of the $\Psi_1$ component 
along the $x^1$-axis 
shows the coincident two kinks, {\it i.e.}, a dark soliton.
The dot in the center denotes 
the point $(\Psi_1,\Psi_2)=0$,  
which corresponds to a singularity 
in the NL$\sigma$M approximation ($\rho$ = const.) 
in (b). 
(b) The isosurface of $s_3=0$ of 
an approximate solution in Eq. (\ref{eq:wall-anti-wall1}) 
with a domain wall and 
an anti-domain wall stretched by a vortex in 
the $O(3)$ NL$\sigma$M, where $M=1$, 
$x_1^1=-3$, $x_2^1=3$, $\phi_1=0$, $\phi_2=\pi$. 
(c) The pseudo-spin texture of 
an approximate solution in Eq.~(\ref{eq:wall-anti-wall1}).
\label{fig:brane-anti-brane} 
} 
\end{figure*}
%%%%%%%%%%%%%%%%%%%%%%%

%%%%%%%%%%%%%%%%%%%%%
\begin{figure}
\begin{center}
\includegraphics[width=0.6\linewidth,keepaspectratio]{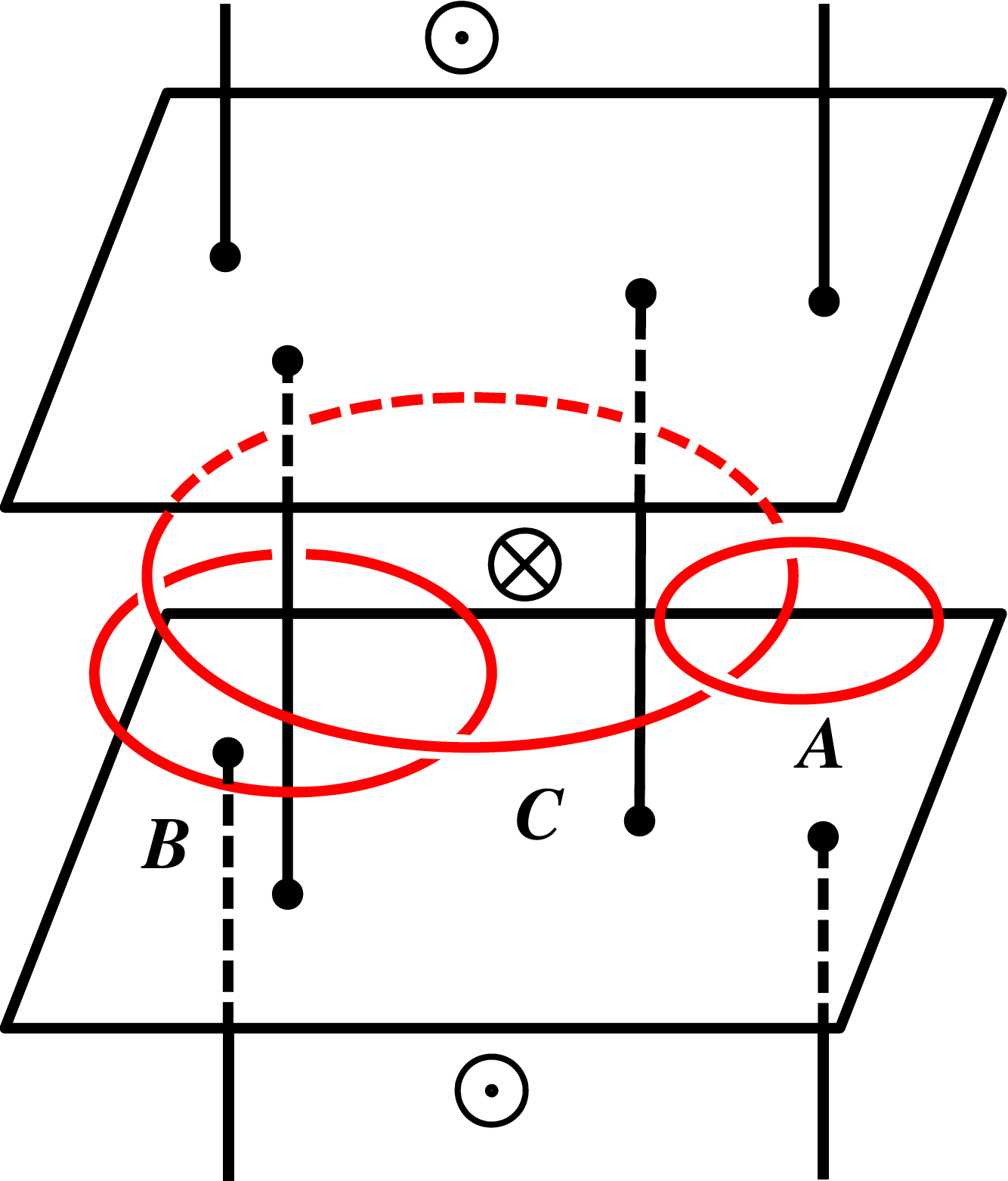} \\
\end{center}
\caption{(Color online)
Loops in the wall-vortex systems. 
While the loop A yields an untwisted vortex ring 
in Fig.~\ref{fig:vorton}(a), 
the loop B (C) yields a vorton, {\it i.e.}, 
a vortex ring twisted once (twice).
A vorton with twisted once 
is shown in Fig.~\ref{fig:vorton}(b).
\label{fig:loops} 
} 
\end{figure}
%%%%%%%%%%%%%%%%%%%%%%%

%%%%%%%%%%%%%%%%%%%%%
\begin{figure*}[t]
\begin{center}
\includegraphics[width=1\linewidth,keepaspectratio]{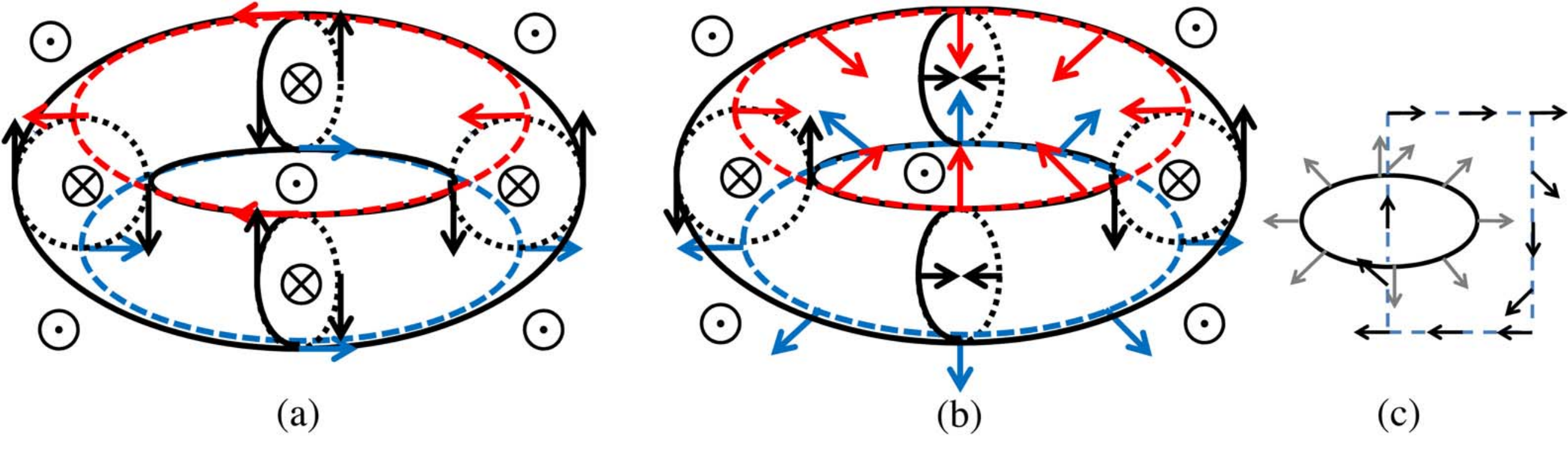}
\end{center}
\caption{(Color online) (a) The pseudo-spin texture of 
an untwisted vortex-ring, and  
(b,c) the pseudo-spin and phase textures of a twisted 
vortex ring, {\it i.e.}, a vorton,  
after the brane-anti-brane annihilation.
(b) The pseudo-spin structure of the vorton.
The torus divides the regions of $\Psi_1$ and $\Psi_2$ 
which repel each other: 
$\Psi_1$ ($\Psi_2$) are filled outside (inside) the torus. 
The vertical section of the torus by the $x^1$-$x^2$ plane  
is a pair of 2D skyrmions and anti-skyrmions. 
While they rotate along the $x^1$-axis 
their pseudo-spins are twisted. 
This spin texture is equivalent to that of a knot 
\cite{Kawaguchi:2008xi,Babaev:2001zy,Babaev:2008zd}.
(c) The phase structure of the vorton. 
The arrows denote the phase of $\Psi_1$ and $\Psi_2$.
The circle denotes the core of vorton where $\Psi_2$ is filled 
and $\Psi_1$ is zero. 
The phase of $\Psi_2$ winds once along that circle. 
The square of the dotted line denotes a loop 
where $\Psi_2$ is zero. 
The phase of $\Psi_1$ winds once along that loop. 
Note that the loops of $\Psi_1$ and $\Psi_2$ are zero respectively 
make a link.
Along the zeros of $\Psi_1$ ($\Psi_2$), 
the phase of $\Psi_2$ ($\Psi_1$) winds once.
\label{fig:vorton} 
} 
\end{figure*}
%%%%%%%%%%%%%%%%%%%%%

We are ready to study a pair of a domain wall and 
an anti-domain wall stretched by vortices. 
An approximate analytic solution of the domain wall pair 
stretched by one vortex, which is 
schematically shown in Fig.~\ref{fig:brane-anti-brane}(a), 
can be given in the $O(3)$ NL$\sigma$M as
\begin{eqnarray}
&& u (x^1,z)= u_{\rm w}(x^1) u_{\rm v}(z), \label{eq:wall-anti-wall1}\\
&& u_{\rm w}(x^1)= e^{-M(x^1 -x^1_1) - i \phi_1} 
 + e^{M(x^1 -x^1_2) - i \phi_2} ,\quad \label{eq:wall-anti-wall2}\\
&& u_{\rm v}(z)= {1 / z}. \label{eq:wall-anti-wall3}
\end{eqnarray}
Here, $x^1_1$ and $x^1_2$ ($x_1^1 < x_2^1$) represent the positions of 
the wall and anti-wall, respectively, 
while $\phi_1$ and $\phi_2$ denote 
the phase of the wall and anti-wall, respectively. 
This solution is good when the distance 
$|x^1_1-x^2_2|$ between the walls is large compared with the mass scale 
$M^{-1}$. 
For our purpose, the phases 
are taken as $\phi_1=\phi_2+\pi$, which means that the $\Psi_1$ component 
has a dark soliton when the intermediate $\Psi_2$ component vanishes.
The isosurface of $s_3=0$ and 
the pseudospin structure of this configuration 
are plotted in Fig.~\ref{fig:brane-anti-brane}(b) and (c), respectively.
In order to avoid the logarithmic bending of the walls,  
one can use $u_{\rm v}(z)$ in Eq.~(\ref{wallvortexcomp}) 
with $N_{v_1}=N_{v_2}$ 
instead of Eq.~(\ref{eq:wall-anti-wall3}), 
as in Fig.~\ref{fig:loops}.
The solution in Eqs.~(\ref{eq:wall-anti-wall1})--(\ref{eq:wall-anti-wall3}) 
of the $O(3)$ NL$\sigma$M has a singularity 
at the midpoint of the vortex stretching the domain walls,
as in Fig.~\ref{fig:brane-anti-brane}.
It is, however, merely an artifact in the NL$\sigma$M approximation
of $\rho$ = const.; 
the singularity does not exist 
in the original theory without such the approximation,
because $\rho$ varies and merely vanishes at that point.

Now let us discuss the dynamics of the wall--anti-wall configuration. 
As in the case without a stretched string,
the configuration itself is unstable to decay, and vortex-loops are created in the $\Psi_1$ component. Since the $\Psi_2$ component is localized along the vortex core, the $s_3=0$ surface forms a torus (ring), where the region of $s_3>0$ is outside the torus whereas the region $s_3<0$ is inside it. 
As the phase of $\Psi_2$ component inside the ring is concerned, the vortex loops are classified into 1) the untwisted case [see Fig.~\ref{fig:vorton}(a)] and 2) the twisted case [see Fig.~\ref{fig:vorton}(b)]. 

1)
If the closed vortex-loop encloses no stretched vortices 
as the loop A in Fig.~\ref{fig:loops}, 
the vortex-loop is not twisted, as in Fig.~\ref{fig:vorton}(a).
Equivalently, the phase of the $\Psi_2$ component 
inside the ring is not wound. 

2)
However, if the vortex-loop encloses $n$ stretched vortices  
as the loops B and C in Fig.~\ref{fig:loops}, 
the vortex-loop is twisted $n$ times. 
It implies that the phase of the $\Psi_2$ component 
inside the ring is wound $n$ times. 
A vortex-loop twisted once, 
which is nothing but a vorton with the minimum twist, 
is shown in Fig.~\ref{fig:vorton}(b). 
The vertical section of the torus by the $x^1$-$x^2$ plane  
is a pair of a skyrmion (coreless vortex) 
and an anti-skyrmion (coreless vortex). 
Moreover, the presence of the stretched vortex
implies that the phase winds anti-clockwise along the loops, 
as can be seen by the arrows on the top and the bottom of the torus 
in Fig.~\ref{fig:vorton}(b). 
When the 2D skyrmion pair rotate along the $x^1$-axis 
their phases are twisted and connected to each other 
at the $\pi$ rotation. 
Note that the zeros of $\Psi_1$ and $\Psi_2$ make a link.
Along the zeros of $\Psi_1$ ($\Psi_2$), 
the phase of $\Psi_2$ ($\Psi_1$) winds once. 
The configuration is nothing but a vorton.

It may be interesting to point out that 
this spin texture is equivalent to the one of 
a knot soliton \cite{Kawaguchi:2008xi,Babaev:2001zy,Babaev:2008zd}, 
{\it i.e.} a topologically nontrivial texture 
with a Hopf charge $\pi_3(S^2) \simeq {\bf Z}$ in an $O(3)$ NL$\sigma$M. 
Mathematically, this fact implies that a vorton is Hopf fibered over a knot.

\begin{figure*}
\begin{center}
\includegraphics[width=1\linewidth,keepaspectratio]{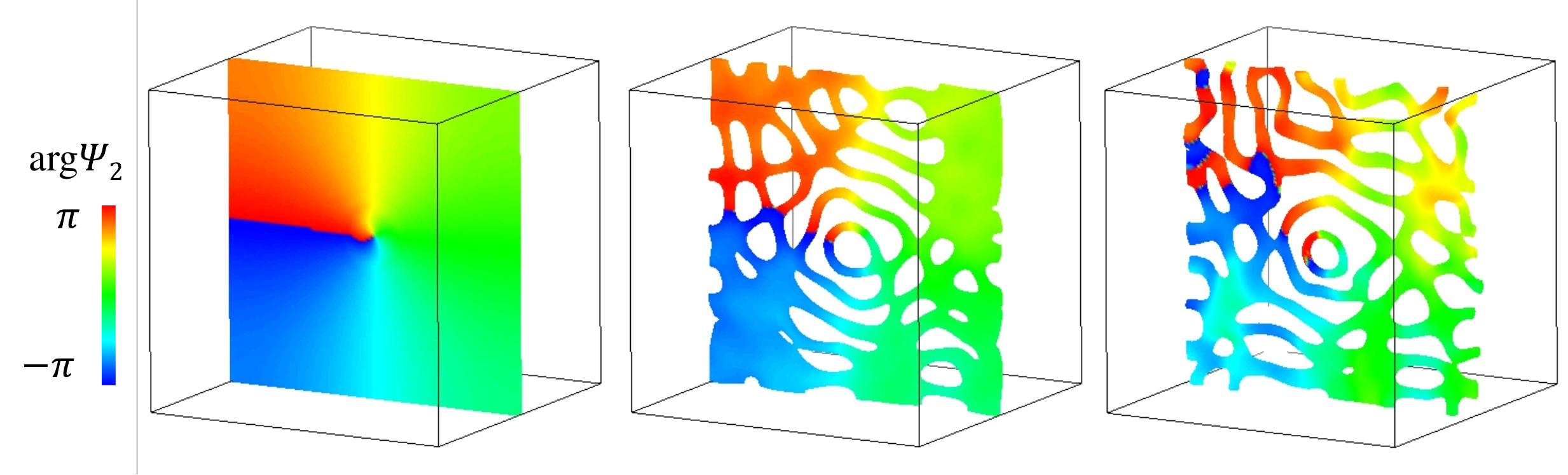}\\
\hspace{1cm}(a)\hspace{5cm}(b)\hspace{5cm}(c)
\end{center}
\caption{(Color online) A numerical simulation of a vorton creation. 
Surfaces are defined by $n_1 - n_2 = 0.18 (\mu_1/g)$ 
($\mu_1=g=1, \hbar = m=1$),  
while color represents the phase of the $\Psi_2$ component.
(a) First, we prepare a pair of a domain wall and an anti-domain wall 
at coincident limit with a vortex winding 
in the $\Psi_2$ component. 
(b) Holes are created in the wall annihilation.
(c) A vorton is created in the center. 
\label{fig:vorton2} 
} 
\end{figure*}

Finally, to confirm a vorton creation of 
a domain wall pair annihilations, 
we show a numerical simulation 
of the time-dependent GP equation $i \hbar \partial_t \Psi_{j} = \delta E/\delta \Psi_j^{\ast}$
for the domain wall pair with a stretched vortex
in Fig.~\ref{fig:vorton2}.
 The numerical scheme to solve the GP equation is a Crank--Nicholson method with the Neumann boundary condition in a cubic box without external potentials.
The box size is $52.1\xi\times 52.1\xi\times 52.1\xi$ with $\xi=\hbar/\sqrt{m\mu_1}$.
We prepare a pair of a domain wall and an anti-domain wall 
at coincident limit with a vortex winding 
in the $\Psi_2$ component. 
Here, for simplicity we put a cylindrically symmetric 
perturbation, which is expected to be induced 
from varicose modes of the string.
Several holes grow after being created, and there appear vortex loops.
Although the holes appear asymmetrically because of the cubic boundary,
the boundary effect is small in the center region and 
the initial perturbation causes a vortex loop there.
The vortex loop enclosing the $\Psi_2$ winding,  
which is nothing but a vorton, is created in the center 
of Fig.~\ref{fig:vorton2}(c).
%%%%%%%%%%%%%%%%%%%%%
%%%%%%%%%%%%%%%%%%%%%

\subsection{Equivalence of the vorton to the three-dimensional skyrmion}

It has been already shown in \cite{Ruostekoski:2001fc,Battye:2001ec} 
that 3D skyrmions are topologically equivalent to vortons 
in two-component BECs. 
In this subsection, we show it in our context of 
the brane-anti-brane annihilations. 

\begin{figure}[t]
\begin{center}
\includegraphics[width=1\linewidth,keepaspectratio]{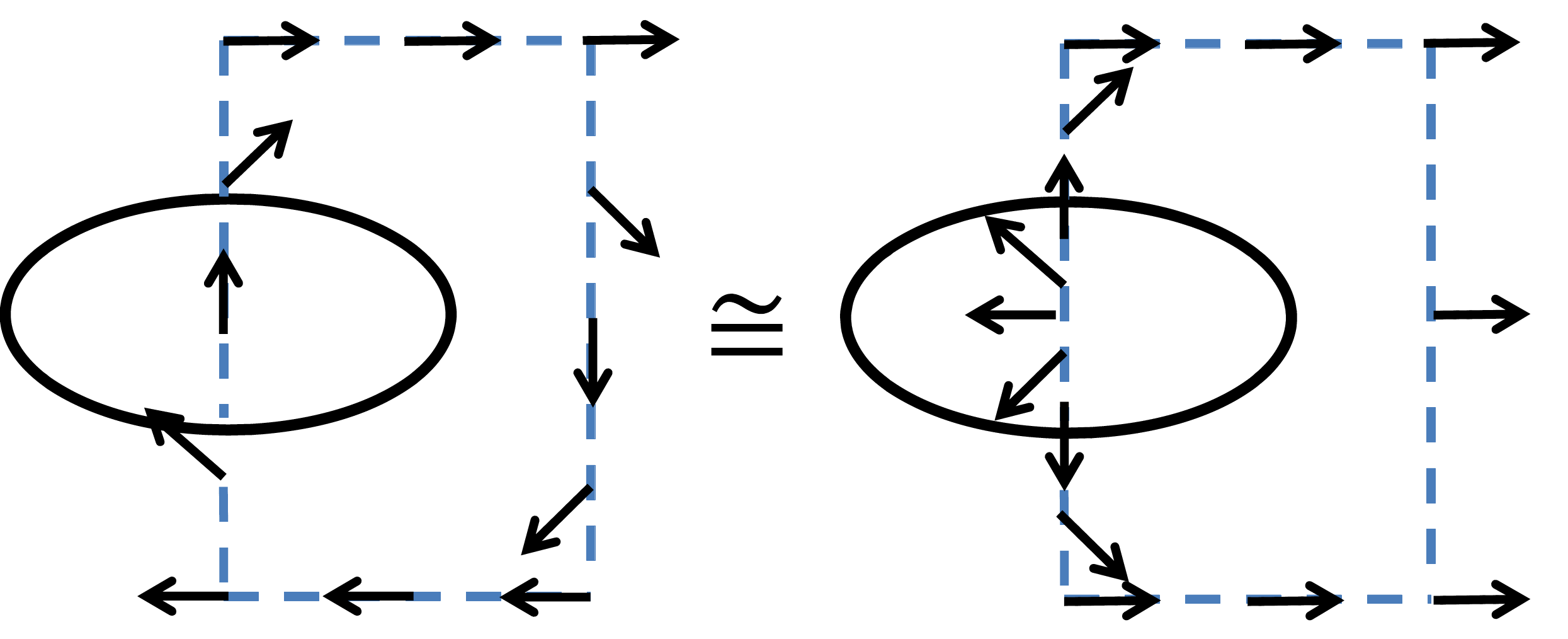}
\end{center}
\caption{(Color online) The equivalence between the vorton and the 3D skyrmion.
 The arrows denote 
the phase of $\Psi_1$ along a large loop 
(of the square of the dotted line) 
going to the boundary 
where  $\Psi_2$ is zero,  
making a link with the vorton core. 
The left panel represents the configuration of 
the vorton from the brane-anti-brane annihilation 
[see Fig.~\ref{fig:vorton}-(a)], 
while the right panel represents the configuration 
of a 3D skyrmion. 
They are topologically isomorphic to each other.
} 
\label{fig:equivalence} 
\end{figure}

In Fig.~\ref{fig:equivalence}, 
the arrows denote 
the phase of $\Psi_1$ along a large loop 
(of the square of the dotted line) 
going to the boundary 
where  $\Psi_2$ is zero,  
making a link with the vorton core. 
The left panel of Fig.~\ref{fig:equivalence} 
represents the phases of $\Psi_1$ and $\Psi_2$ of 
the vorton from the brane-anti-brane annihilation 
[see also Fig.~\ref{fig:vorton}(b)].
This is topologically equivalent to  
the right panel of Fig.~\ref{fig:equivalence}.
Here we show that the phase structure of the right panel 
is that of a 3D skyrmion. 
They are topologically isomorphic to each other.
%%%%%%%%%%%%%%%%%%%%%
%%%%%%%%%%%%%%%%%%%%%

First, let us introduce the matrix $U$ as
\begin{eqnarray}
\left(\begin{array}{c}
\Psi_1\\
\Psi_2
\end{array}
\right)
=
\left(\begin{array}{cc}
\Psi_1 & - \Psi_2^*\\
\Psi_2 &   \Psi_1^*
\end{array}
\right)
\left(\begin{array}{c}
1\\
0
\end{array}
\right) 
= U \left(\begin{array}{c}
1\\
0
\end{array}
\right), \label{eq:Psi-3D-skyrmion}
\end{eqnarray}
with 
\begin{eqnarray}
U
\equiv
\left(\begin{array}{cc}
\Psi_1 & - \Psi_2^*\\
\Psi_2 &   \Psi_1^*
\end{array}
\right) \label{eq:SU(2)element}
\end{eqnarray}
being an element of an $SU(2)$ group, when 
\begin{eqnarray}
 \det U = |\Psi_1|^2 + |\Psi_2|^2 =1. \label{eq:SU(2)}
\end{eqnarray}

The GP energy functional given in Eq.~(\ref{GPene}) 
is not $SU(2)$ symmetric in general. 
When the relations 
\begin{equation}
 g_{11}=g_{22}=g_{12}, \quad 
 \mu_1= \mu_2  \label{eq:SU(2)symmetric}
\end{equation}
hold, 
the GP energy functional is $SU(2)$ symmetric, 
and Eq.~(\ref{eq:SU(2)}) holds 
(up to overall constant) \cite{footnote1}. 

Even when the GP energy functional given in Eq.~(\ref{GPene}) 
is not $SU(2)$ symmetric, 
we can approximately 
consider the parametrization by $U$ in Eq.~(\ref{eq:SU(2)element}).
A rotationally symmetric configuration 
of a 3D skyrmion can be given by \cite{Skyrme:1961vq}
\begin{eqnarray}
U = \exp i \left( f(r){{\bf r} \over |{\bf r}|} \cdot {\bf \sigma} \right)
 \label{eq:3d-skyrmion}
\end{eqnarray}
with a function $f(r)$ with the boundary condition 
\begin{eqnarray} 
 f(r=0) = n \pi, \quad f(r=R) = 0,
\end{eqnarray}
where $R$ is the size of the system.
Here, $n \in {\bf Z}$ is an element of 
the third homotopy group
$\pi_3 (SU(2)) \simeq {\bf Z}$.

In the polar coordinates $(r,\theta,\phi)$, 
\begin{equation}
{{\bf r} \over |{\bf r}|}= 
(\sin \theta \cos \phi,  \sin \theta \sin \phi, \cos \theta) .
\end{equation}
By using the formula, 
$\exp(i \Theta {\bf n}\cdot {\bf \sigma})
=\cos \Theta + i {\bf n}\cdot {\bf \sigma} \sin \Theta$ 
for ${\bf n}^2=1$,  
the 3D skyrmion in 
Eq.~(\ref{eq:Psi-3D-skyrmion}) with 
$U$ in Eq.~(\ref{eq:3d-skyrmion}) can be obtained as
\begin{eqnarray}
\left(\begin{array}{c}
\Psi_1\\
\Psi_2
\end{array}
\right)
= \left(\begin{array}{c}
\cos f(r) - i \sin f(r) \cos \theta \\
\sin f(r) \sin \theta e^{-i\phi}
\end{array}
\right).\label{eq:3d-skyrmion2}
\end{eqnarray}

First, let us study the phase structure of $\Psi_1$ 
of the 3D skyrmion in Eq.~(\ref{eq:3d-skyrmion2}).
At the boundaries and the origin, Eq.~(\ref{eq:3d-skyrmion2}) becomes
\begin{eqnarray}
 \left(\begin{array}{c}
\Psi_1\\
\Psi_2
\end{array}\right) 
&=& \left(\begin{array}{c}
1\\
0
\end{array}\right)  \mbox{ at } r= R ,\nonumber\\
\left(\begin{array}{c}
\Psi_1\\
\Psi_2
\end{array}\right) 
&=& \left(\begin{array}{c}
(-1)^n\\
0
\end{array}\right)  \mbox{ at } r= 0 .
\label{eq:r=R,0}
\end{eqnarray}
Along the $x^1$-axis ($\theta =0,\pi$), Eq.~(\ref{eq:3d-skyrmion2}) becomes 
\begin{eqnarray}
 \left(\begin{array}{c}
\Psi_1\\
\Psi_2
\end{array}\right) 
&=& \left(\begin{array}{c}
\exp [-if(r)]\\
0
\end{array}\right) \mbox{ at }  \theta =0 ,\nonumber\\
\left(\begin{array}{c}
\Psi_1\\
\Psi_2
\end{array}\right) 
&=& \left(\begin{array}{c}
\exp [if(r)]\\
0
\end{array}\right) \mbox{ at }  \theta =\pi , 
\label{eq:Psi1}
\end{eqnarray}
Eqs.~(\ref{eq:r=R,0}) and (\ref{eq:Psi1}) 
show, in the case of $n=1$, the phase structure of $\Psi_1$
in the right panel of 
Fig.~\ref{fig:equivalence}. 

Second, let us study the phase structure of $\Psi_2$. 
We consider the ring defined by 
$\theta=\pi/2$ and the radius $r$ such that $f(r)=\pi/2$.
Along this ring, the $\Psi_i$ are 
\begin{eqnarray}
\left(\begin{array}{c}
\Psi_1\\
\Psi_2
\end{array}
\right)
= \left(\begin{array}{c}
 0 \\
 e^{-i\phi}
\end{array}
\right) .
\end{eqnarray}
The $\Psi_2$ component winds once along this ring, 
as in Fig.~\ref{fig:vorton}(c).
This winding of the $\Psi_2$ component originates 
from the winding in the brane-anti-brane configuration 
in Eq.~(\ref{eq:wall-anti-wall3}).

We thus have seen that the 3D skyrmion in 
Eq.~(\ref{eq:Psi-3D-skyrmion}) is 
topologically equivalent to a vorton.

%%%%%%%%%%%%%%%%%%%%%%%%%%%%%%%%%%%%%%%
\section{Summary and discussion \label{sec:summary}}
We have studied a mechanism to create a vorton 
or three dimensional skyrmion in 
phase separated two-component BECs. 
We consider a pair of a domain wall and an anti-domain wall 
with vortices stretched between them. 
The $\Psi_2$ component is sandwiched by 
the regions of the $\Psi_1$ component, 
where the phase difference of $\Psi_1$'s 
in the two separated regions 
is taken to be $\pi$. 
When the domain wall pair decays, 
there appear vortex loops of the $\Psi_1$ component with 
the $\Psi_2$ component trapped inside their cores.
If a $\Psi_1$ vortex loop encloses one stretched vortex, 
it becomes a vorton. 
More generally, if the vortex loop encloses 
$n$ of the stretched vortices, 
it becomes a vortex ring with the phase of $\Psi_2$ 
twisted $n$ times.
We also have confirmed that the vorton ($n=1$) is topologically 
equivalent to a 3D skyrmion.

Experimentally this can be realized by preparing 
the phase separation in the order $\Psi_1$, $\Psi_2$ and $\Psi_1$ components, 
and rotating the intermediate $\Psi_2$ component.
By selectively removing the filling $\Psi_2$ component  
with a resonant laser beam, the collision of the brane and anti-brane 
can be made, to create vortons.

Once created in the laboratory, one can study 
the stability and dynamics of a vorton experimentally. 
The vorton will propagate along the direction perpendicular to the initial configuration of the branes. Therefore, to investigate the dynamics of a vorton, we need to prepare a large size of cloud in that direction. In the case of untwisted vortex loop (usual loop, not a vorton), it will easily shrink and eventually decay into phonons if the thermal dissipation works enough. However, the vorton should be stable against the shrinkage and will propagate to reach the surface of the atomic cloud. Such a difference must be a benchmark to detect vortons in experiments.

On the other hand, 
the thermal and quantum fluctuations may make the vorton {\it unstable}. 
Our numerical calculations rely on the mean field GP theory. The topological charge of a vorton is the winding of the phase of $\Psi_2$ along the closed loop (which is proportional to the superflow along the closed loop). Since this topological charge is defined only in the vicinity of the vorton, there is a possibility that it can be unwound, once quantum/thermal fluctuation is taken into account beyond the mean field theory. Quantum mechanically, such a decay is caused by an instanton effect (quantum tunneling). This process also resembles the phase slip of superfluid rings. The vorton decay by the quantum and thermal tunneling is considered to be an important process in high energy physics and cosmology, since it will radiate high energy particles such as photons, which may explain some high energy astrophysical phenomena observed in our Universe. Therefore it would be important that one realizes vortons in laboratory by using ultra-cold atomic gases; it may simulate a vorton decay with emitting phonons quantum mechanically, beyond the mean field approximation.

In this paper, we have mainly 
studied topological aspects of 
the vorton creation using the NL$\sigma$M approximation.
In order to study dynamics of topological defects 
beyond this approximation, 
we need a precise form of the interaction between the defects.
An analytic form of the interaction between vortices 
was derived in the case of miscible ($c_2>0$) two-component BECs, 
and it was applied to the analysis of vortex lattices 
\cite{Eto:2011wp}.
Extension to the case of the immiscible case ($c_2<0$) 
focused in this paper
will be useful to study the interaction 
between vortices attached to domain walls, 
and that between vortons and/or walls 

In our previous paper \cite{Kasamatsu:2010aq}, 
we discussed that 
the domain wall in two-component BECs 
can be regarded as a D2-brane,  
as the D-brane soliton \cite{Gauntlett:2000de,Shifman:2002jm,
Isozumi:2004vg,Eto:2006pg,Eto:2008mf} 
in field theory,  
where ``D$p$-brane" implies a D-brane with $p$ space dimensions.
This is because 
the string endpoints are electrically charged 
under $U(1)$ gauge field of 
the Dirac-Born-Infeld (DBI) action for a D-brane \cite{Dirac:1962iy}.
In our context, the $U(1)$ gauge field is obtained by 
a duality transformation 
from $U(1)$ Nambu-Goldstone mode of the domain wall.  
Since the D-brane soliton \cite{Kasamatsu:2010aq} in two-component BECs, 
precisely coincides with a BIon \cite{Gibbons:1997xz}, 
{\it i.e.}, a soliton solution of the DBI action of a D-brane, 
the domain wall can be regarded as a D2-brane.

On the other hand, it is known in string theory \cite{Sen} that 
when a D$p$-brane and an anti-D$p$-brane annihilate on collision, 
there appear D$(p-2)$ branes. 
If we want to regard our domain wall as a D2-brane, 
the pair annihilation of a D2-brane and anti-D2-brane 
should result in the creation of D0-branes. 
Therefore, a discussion along this line leads us to suggest 
a possible interpretation of 3D skyrmions as D0-branes, 
which are point-like objects.

\acknowledgments
M.~N. would like to thank Michikazu Kobayashi for 
a useful discussion on three dimensional skyrmions.
This work was supported by KAKENHI from JSPS 
(Grant Nos. 21340104, 21740267 and 23740198).
This work was also supported 
by the ``Topological Quantum Phenomena'' 
(Nos. 22103003 and 23103515) 
Grant-in Aid for Scientific Research on Innovative Areas   
from the Ministry of Education, Culture, Sports, Science and Technology 
(MEXT) of Japan.



\begin{thebibliography}{99}


%\cite{Witten:1984eb}
\bibitem{Witten:1984eb} 
  E.~Witten,
  %``Superconducting Strings,''  
Nucl.\ Phys.\ B {\bf 249}, 557 (1985).  %%CITATION = NUPHA,B249,557;%%

%\cite{Davis:1988jq}
\bibitem{Davis:1988jq} 
  R.~L.~Davis and E.~P.~S.~Shellard,
  %``The Physics Of Vortex Superconductivity. 2,''  
Phys.\ Lett.\ B {\bf 209}, 485 (1988).  %%CITATION = PHLTA,B209,485;%%

%\cite{Radu:2008pp}
\bibitem{Radu:2008pp} 
  E.~Radu and M.~S.~Volkov,
  %``Existence of stationary, non-radiating ring solitons in field theory: knots and vortons,''  
Phys.\ Rept.\  {\bf 468}, 101 (2008). 
% [arXiv:0804.1357 [hep-th]].  %%CITATION = ARXIV:0804.1357;%%

\bibitem{Volovik}
G.~E.~Volovik,
{\it The Universe in a Helium Droplet}, 
Clarendon Press,  Oxford (2003).

\bibitem{Vilenkin:2000}
A.~Vilenkin and E.~P.~S.~Shellard, 
{\it Cosmic Strings and Other Topological Defects}, (Cambridge Monographs on Mathematical Physics), Cambridge University Press (July 31, 2000).


%\cite{Skyrme:1961vq}
\bibitem{Skyrme:1961vq} 
  T.~H.~R.~Skyrme,
  %``A Nonlinear field theory,''  
Proc.\ Roy.\ Soc.\ Lond.\ A {\bf 260}, 127 (1961);  %%CITATION = PRSLA,A260,127;%%
%\cite{Skyrme:1962vh}
%\bibitem{Skyrme:1962vh} 
%  T.~H.~R.~Skyrme,
  %``A Unified Field Theory of Mesons and Baryons,'' 
 Nucl.\ Phys.\  {\bf 31}, 556 (1962).  %%CITATION = NUPHA,31,556;%%

%\cite{Manton:2004tk}
\bibitem{Manton:2004tk}
  N.~S.~Manton and P.~Sutcliffe,
  {\it Topological solitons},
%\href{http://www.slac.stanford.edu/spires/find/hep/www?irn=6000355}{SPIRES entry}
{\it  Cambridge, UK: Univ. Pr. (2004) 493 p}

%\cite{Ruostekoski:2001fc}
\bibitem{Ruostekoski:2001fc} 
  J.~Ruostekoski and J.~R.~Anglin,
  %``Creating vortex rings and three-dimensional skyrmions in Bose-Einstein condensates,''  
Phys.\ Rev.\ Lett.\  {\bf 86}, 3934 (2001).
%  [cond-mat/0103310].  %%CITATION = COND-MAT/0103310;%%

%\cite{Battye:2001ec}
\bibitem{Battye:2001ec} 
  R.~A.~Battye, N.~R.~Cooper and P.~M.~Sutcliffe,
  %``Stable skyrmions in two-component Bose-Einstein condensates,''  
Phys.\ Rev.\ Lett.\  {\bf 88}, 080401 (2002).
%  [cond-mat/0109448].  %%CITATION = COND-MAT/0109448;%%

\bibitem{3D-Skyrmions2}
  U.~A.~Khawaja and H.~T.~C.~Stoof,
%``Skyrmions in a ferromagnetic Bose?Einstein condensate,"
Nature (London) {\bf 411}, 918 (2001),  
%\cite{Khawaja:2001zz}
%\bibitem{Khawaja:2001zz}
%  U.~A.~Khawaja and H.~T.~C.~Stoof,
%  ``Skyrmion physics in Bose-Einstein ferromagnets,''
  Phys.\ Rev.\  A {\bf 64}, 043612 (2001).
  %%CITATION = PHRVA,A64,043612;%%

%\cite{Savage:2003hh}
\bibitem{Savage:2003hh} 
  C.~M.~Savage and J.~Ruostekoski,
  %``Energetically stable particle-like skyrmions in a trapped Bose-Einstein condensate,''  
Phys.\ Rev.\ Lett.\  {\bf 91}, 010403 (2003).
%  [cond-mat/0306112].  %%CITATION = COND-MAT/0306112;%%

%\cite{Ruostekoski:2004pj}
\bibitem{Ruostekoski:2004pj} 
  J.~Ruostekoski,
  %``Stable particlelike solitons with multiply-quantized vortex lines in Bose-Einstein condensates,''  
Phys.\ Rev.\ A {\bf 70}, 041601 (2004).
%  [cond-mat/0408376].  %%CITATION = COND-MAT/0408376;%%

\bibitem{Wuster:2005}
S.~Wuster, T.~E.~Argue, and C.~M.~Savage,
%Numerical study of the stability of skyrmions in Bose-Einstein condensates
Phys.\ Rev.\ A {\bf 72}, 043616 (2005).

\bibitem{Oshikawa:2006}
I.~F.~Herbut and M.~Oshikawa, 
%Stable Skyrmions in spinor condensates
Phys.\ Rev.\ Lett.\ {\bf 97}, 080403 (2006);
%arXiv:cond-mat/0604557
%
%Skyrmion in spinor condensates and its stability in trap potentials 
A.~Tokuno, Y.~Mitamura, M.~Oshikawa, I.~F.~Herbut, 
Phys.\ Rev.\ A {\bf 79}, 053626 (2009). 
%arXiv:0812.2736 


%\cite{Metlitski:2003gj}
\bibitem{Metlitski:2003gj} 
  M.~A.~Metlitski and A.~R.~Zhitnitsky,
  %``Vortex rings in two-component Bose-Einstein condensates,''  
JHEP {\bf 0406}, 017 (2004).
%  [cond-mat/0307559].  %%CITATION = COND-MAT/0307559;%%

%\cite{Bedaque:2011sb}
\bibitem{Bedaque:2011sb}
  P.~F.~Bedaque, E.~Berkowitz and S.~Sen,
  %``Stable vortex loops in two-species BECs,''
  arXiv:1111.4507 [cond-mat.quant-gas].
  %%CITATION = ARXIV:1111.4507;%%

\bibitem{Leslie:2009}
L.~S.~Leslie, A.~Hansen, K.~C.~Wright, B.~M.~Deutsch, and N.~P.~Bigelow, 
Phys.\ Rev.\ Lett.\ {\bf 103}, 250401 (2009);
J.~Choi, W.~J.~Kwon, and Y.~Shin 
Phys.\ Rev.\ Lett.\ {\bf 108}, 035301 (2012). 

\bibitem{Stoof} 
H.~T.~C.~Stoof, E.~Vliegen, and U.~Al~Khawaja,  
%{\it Monopoles in an Antiferromagnetic Bose-Einstein Condensate.}
Phys. Rev. Lett. {\bf 87}, 120407 (2001).

\bibitem{Martikainen} 
J.~-P.~Martikainen, A.~Collin, and K.~-A.~Suominen,  
%{\it Creation of a Monopole in a Spinor Condensate.}
Phys. Rev. Lett. {\bf 88}, 090404 (2002). 

\bibitem{Savage}
C.~M.~Savage and J.~Ruostekoski, 
%{\it Dirac monopoles and dipoles in ferromagnetic spinor Bose-Einstein condensates.}
Phys. Rev. A {\bf 68}, 043604 (2003). 

\bibitem{Pietila:2009}
V.~Pietil\"a and M.~M\"ott\"onen, 
Phys.\ Rev.\ Lett.\ {\bf 103}, 030401 (2009).

%\cite{Kawaguchi:2008xi}
\bibitem{Kawaguchi:2008xi} 
  Y.~Kawaguchi, M.~Nitta and M.~Ueda,
  %``Knots in a Spinor Bose-Einstein Condensate,''  
Phys.\ Rev.\ Lett.\  {\bf 100}, 180403 (2008). 
% [Erratum-ibid.\  {\bf 101}, 029902 (2008)]  [arXiv:0802.1968 [cond-mat.other]].  %%CITATION = ARXIV:0802.1968;%%

%\cite{Semenoff:2006vv}
\bibitem{Semenoff:2006vv}
  G.~W.~Semenoff and F.~Zhou,
  %``Discrete symmetries and 1/3-quantum vortices in condensates of F=2 cold
  %atoms,''
  Phys.\ Rev.\ Lett.\  {\bf 98}, 100401 (2007);
%  [arXiv:cond-mat/0610162].
  %%CITATION = PRLTA,98,100401;%%
%\cite{Kobayashi:2008pk}
%\bibitem{Kobayashi:2008pk}
  M.~Kobayashi, Y.~Kawaguchi, M.~Nitta and M.~Ueda,
  %``Collision Dynamics and Rung Formation of Non-Abelian Vortices,''
  Phys.\ Rev.\ Lett.\  {\bf 103}, 115301 (2009).
%  [arXiv:0810.5441 [cond-mat.other]].
  %%CITATION = PRLTA,103,115301;%%

\bibitem{review} 
K. Kasamatsu, M. Tsubota and M. Ueda, 
%``Vortices in Multicomponet Bose-Einstein Condensates,"
Int. J. Mod. Phys. B {\bf 19}, 1835 (2005);
%\cite{Kawaguchi:2010mu}
%\bibitem{Kawaguchi:2010mu} 
  Y.~Kawaguchi, M.~Kobayashi, M.~Nitta and M.~Ueda,
  %``Topological Excitations in Spinor Bose-Einstein Condensates,''  
Prog.\ Theor.\ Phys.\ Suppl.\  {\bf 186}, 455 (2010)  
%[arXiv:1006.5839 [cond-mat.quant-gas]].  %%CITATION = ARXIV:1006.5839;%%
%CITATION = ARXIV:1006.5839;%%
M.~Ueda and Y.~Kawaguchi,
%``Spinor Bose-Einstein condensates,"
arXiv:1001.2072 (2010).

\bibitem{Thalhammer}
G. Thalhammer, {\it et al.} 
%{\it Double Species Bose-Einstein Condensate with Tunable Interspecies Interactions.}
Phys. Rev. Lett. {\bf 100}, 210402 (2008).
\bibitem{Papp}
S. B. Papp, J. M. Pino, and C. E. Wieman, 
%{\it Tunable Miscibility in a Dual-Species Bose-Einstein Condensate.}
Phys. Rev. Lett. {\bf 101}, 040402 (2008). 

\bibitem{Kasamatsu2} 
K. Kasamatsu, M. Tsubota, and M. Ueda,  
%{\it Spin textures in rotating two-component Bose-Einstein condensates.}
Phys. Rev. A {\bf 71}, 043611 (2005). 

%\cite{Babaev:2001zy}
\bibitem{Babaev:2001zy} 
  E.~Babaev, L.~D.~Faddeev and A.~J.~Niemi,
  %``Hidden symmetry and knot solitons in a charged two-condensate Bose system,''  
Phys.\ Rev.\ B {\bf 65}, 100512 (2002).
%  [cond-mat/0106152 [cond-mat.supr-con]].  %%CITATION = COND-MAT/0106152;%%

%\cite{Babaev:2008zd}
\bibitem{Babaev:2008zd} 
  E.~Babaev,
  %``Non-Meissner electrodynamics and knotted solitons in two-component superconductors,''  
Phys.\ Rev.\ B {\bf 79}, 104506 (2009).
%  [arXiv:0809.4468 [cond-mat.supr-con]].  %%CITATION = ARXIV:0809.4468;%%

%\cite{Gauntlett:2000de}
\bibitem{Gauntlett:2000de} 
  J.~P.~Gauntlett, R.~Portugues, D.~Tong and P.~K.~Townsend,
  %``D-brane solitons in supersymmetric sigma models,''  
Phys.\ Rev.\ D {\bf 63}, 085002 (2001).
%  [hep-th/0008221].  %%CITATION = HEP-TH/0008221;%%

%\cite{Shifman:2002jm}
\bibitem{Shifman:2002jm} 
  M.~Shifman and A.~Yung,
  %``Domain walls and flux tubes in N=2 SQCD: D-brane prototypes,''  
Phys.\ Rev.\ D {\bf 67}, 125007 (2003).
%  [hep-th/0212293].  %%CITATION = HEP-TH/0212293;%%

%\cite{Isozumi:2004vg}
\bibitem{Isozumi:2004vg} 
  Y.~Isozumi, M.~Nitta, K.~Ohashi and N.~Sakai,
  %``All exact solutions of a 1/4 Bogomol'nyi-Prasad-Sommerfield equation,''  Phys.\ 
Rev.\ D {\bf 71}, 065018 (2005).
%  [hep-th/0405129].  %%CITATION = HEP-TH/0405129;%%

%\cite{Eto:2006pg}
\bibitem{Eto:2006pg} 
  M.~Eto, Y.~Isozumi, M.~Nitta, K.~Ohashi and N.~Sakai,
  %``Solitons in the Higgs phase: The Moduli matrix approach,''  
J.\ Phys.\ A A {\bf 39}, R315 (2006).
%  [hep-th/0602170].  %%CITATION = HEP-TH/0602170;%%

%\cite{Eto:2008mf}
\bibitem{Eto:2008mf} 
  M.~Eto, T.~Fujimori, T.~Nagashima, M.~Nitta, K.~Ohashi and N.~Sakai,
  %``Dynamics of Strings between Walls,''  
Phys.\ Rev.\ D {\bf 79}, 045015 (2009).
%  [arXiv:0810.3495 [hep-th]].  %%CITATION = ARXIV:0810.3495;%%

\bibitem{Polchinski:1995mt}
 J. Polchinski,
%{\it Dirichlet Branes and Ramond-Ramond Charges.} 
Phys. Rev. Lett. {\bf 75}, 4724 (1995). 
\bibitem{Leigh}
R. G. Leigh, 
%{\it Dirac-Born-Infeld Action from Dirichlet Sigma Model. }
Mod. Phys. Lett.  {\bf A4}, 2767 (1989). 

\bibitem{Polchinskibook}
J. Polchinski, 
{\it String Theory} (Cambridge Univ. Press, Cambridge, 1998). 

%\cite{Kasamatsu:2010aq}
\bibitem{Kasamatsu:2010aq} 
  K.~Kasamatsu, H.~Takeuchi, M.~Nitta and M.~Tsubota,
  %``Analogues of D-branes in Bose-Einstein condensates,''  
JHEP {\bf 1011}, 068 (2010). 
% [arXiv:1002.4265 [cond-mat.quant-gas]].  %%CITATION = ARXIV:1002.4265;%%

%\cite{Borgh:2012es}
\bibitem{Borgh:2012es} 
  M.~O.~Borgh and J.~Ruostekoski,
  %``Topological interface engineering and defect crossing in ultracold atomic gases,''  
arXiv:1202.5679 [cond-mat.quant-gas].  %%CITATION = ARXIV:1202.5679;%%

\bibitem{Bradley}
D. I. Bradley, S. N. Fisher, A. M. Gu\'enault, R. P. Haley, J. Kopu, H. Martin, 
G. R. Pickett, J. E. Roberts and V. Tsepelin, 
%{\it Relic topological defects from brane annihilation simulated in superfluid $^3$He.}
Nature Phys. {\bf 4}, 46 (2008).

\bibitem{Anderson}
B. P. Anderson, P. C. Haljan, C. A. Regal, D. L. Feder, L. A. Collins, C. W. Clark, and E. A. Cornell, 
%{\it Watching Dark Solitons Decay into Vortex Rings in a Bose-Einstein Condensate. }
Phys. Rev. Lett. {\bf 86}, 2926 (2001). 

\bibitem{Takeuchi:2011}
%``Vortex Formations from Domain Wall Annihilations in Two-Component Bose-Einstein Condensates"
H.~Takeuchi, K.~Kasamatsu, M.~Nitta and M.~Tsubota, 
J.\ Low Temp.\ Phys.\ {\bf 162}, 243 (2011);
%\cite{Takeuchi:2012ee}
%\bibitem{Takeuchi:2012ee} 
  H.~Takeuchi, K.~Kasamatsu, M.~Tsubota and M.~Nitta,
  %``Tachyon Condensation in Bose--Einstein Condensates,''  
arXiv:1205.2330 [cond-mat.quant-gas].  %%CITATION = ARXIV:1205.2330;%%

\bibitem{Pethickbook}
C.~J.~Pethick and H.~Smith,  
{\it Bose-Einstein Condensation in Dilute Gases, 2nd ed.} (Cambridge Univ. Press, Cambridge, 2008).

\bibitem{Sen}
A. Sen, 
%{\it Tachyon dynamics in open string theory.}
Int. J. Mod. Phys. A {\bf 20}, 5513 (2005). 
%\cite{Hashimoto:2001rj}

\bibitem{AndersonToulouse}
P. W. Anderson and G. Toulouse, 
%{\it Phase Slippage without Vortex Cores: Vortex Textures in Superfluid $^3$He.}
Phys. Rev. Lett. {\bf 38}, 508 (1977).

%\cite{Polyakov:1975yp}
\bibitem{Polyakov:1975yp} 
  A.~M.~Polyakov and A.~A.~Belavin,
  %``Metastable States of Two-Dimensional Isotropic Ferromagnets,''  
JETP Lett.\  {\bf 22}, 245 (1975)  [Pisma Zh.\ Eksp.\ Teor.\ Fiz.\  {\bf 22}, 503 (1975)].  %%CITATION = JTPLA,22,245;%%


\bibitem{footnote1}
In the $SU(2)$ symmetric case, 
the conditions in Eq.~(\ref{eq:SU(2)symmetric}) imply 
%\begin{equation} 
  $c_1=c_2=0$    \label{eq:SU(2)symmetric2}
%\end{equation}
in the $O(3)$ NL$\sigma$M description, 
and consequently there is no potential. 
By noting the isomorphism $SU(2)\simeq S^3$,
the target space $S^2$ is a quotient 
of $S^3$ by gauge symmetry: $S^2 = S^3/U(1)$, 
which is nothing but the Hopf fiberation.

%\cite{Eto:2011wp}
\bibitem{Eto:2011wp} 
  M.~Eto, K.~Kasamatsu, M.~Nitta, H.~Takeuchi and M.~Tsubota,
  %``Interaction of half-quantized vortices in two-component Bose-Einstein condensates,''  
Phys.\ Rev.\ A {\bf 83}, 063603 (2011);
%  [arXiv:1103.6144 [cond-mat.quant-gas]].  %%CITATION = ARXIV:1103.6144;%%
P.~Mason and A.~Aftalion, 
Phys.\ Rev.\ A {\bf 84}, 033611 (2011); 
A.~Aftalion, P.~Mason, and J.~Wei, 
Phys.\ Rev.\ A {\bf 85}, 033614 (2012).


\bibitem{Dirac:1962iy}
P. A. M. Dirac, 
%{\it An Extensible model of the electron.}
Proc. Roy. Soc. Lond. A {\bf 268}, 57 (1962); 
%\bibitem{Born:1934gh}
M. Born and L. Infeld, 
%{\it Foundations of the New Field Theory.}
Proc. Roy. Soc. Lond. A {\bf 144}, 425 (1934). 

\bibitem{Gibbons:1997xz}
G. W. Gibbons, 
%{\it Born-Infeld particles and Dirichlet p-branes.}
Nucl. Phys. B {\bf 514}, 603 (1998). 
%\bibitem{Callan}
C. G. Callan and J. M. Maldacena, 
%{\it Brane dynamics from the Born-Infeld action.}
Nucl. Phys. B {\bf 513}, 198 (1998).


\end{thebibliography}
\end{document}